\documentclass[10pt]{ufn}

\usepackage{psfig}

\begin{document}

\sloppypar

\title{\bf Sky surveys and deep fields of ground-based \\ 
and space telescopes}

\author{V.P.\,Reshetnikov}

\institute{Astronomical Institute of St.Petersburg State University,
Universitetskii pr. 28, Petrodvoretz, 198504 Russia; resh@astro.spbu.ru}

\authorrunning{V.Reshetnikov}
\titlerunning{Sky surveys and deep fields}

\abstract{Selected results obtained in major observational sky
surveys (DSS, 2MASS, 2dF, SDSS) and deep field observations
(HDF, GOODS, HUDF, etc.) are reviewed. Modern surveys provide
information on the characteristics and space distribution of
millions of galaxies. Deep fields allow one to study galaxies at the
stage of formation and to trace their evolution over billions of
years. The wealth of observational data is altering the face of
modern astronomy: the formulation of problems and their solutions 
are changing and all the previous knowledge, from planetary studies
in the solar system to the most distant galaxies and quasars, is
being revised.
}
\titlerunning{Sky surveys and deep fields}
\maketitle

\section{Introduction}

People have always tried to understand the surrounding world.
By now, many figures and plans have been collected showing
the way people have tried to systematize their knowledge,
from geographic maps to pure speculative scheme of the Universe
([1], Fig. 1).

For a very long time, humankind included numerous ``fixed'' stars,
planets, the Sun, and the Moon into its sphere of interests.
Probably, the first systematic survey of all that is visible
by the naked eye was performed by Hipparchus in the 2nd century BC.
He drew up a catalog including about 850 stars. After almost two
thousand years, at the end of the 18th century, French astronomer
Charles Messier published the first catalog including not stars
but stellar clusters and nebulae. As we know know, about a third
of these nebulae are extragalactic bodies -- external galaxies.
However, Messier was not interested in the dim fuzzy spots he
discovered. He was primarily interested in comets and compiled
this catalog in order to distinguish comets from fixed nebulae.

William Herschel (1738-1822, ``Coelorum perrupit claustra'')
was the first to formulate the problem of global sky surveys
to study the structure and evolution of the world outside the
solar system [2]. To survey stars in the sky, he applied the
original method of ``star gauging'' (counting the number of
stars in selected sky areas\footnote{This was one of a few cases
where an astronomer, in full agreement with the commonplace
opinion about his kind of work, actually ``counted stars''
by looking through a telescope.}) and statistical data analysis.
This allowed him to establish the general shape of our Galaxy
and to estimate correctly its oblatness ($\sim$1/5). 
Another great merit of Herschel was the first systematic survey
of faint nebulae and an attempt to establish regularities in their
large-scale distribution. He discovered more
than 2.5 thousand nebulae and star clusters, of which 80\% are
other galaxies. Herschel was first to attempt to estimate the
size of dim nebulae and to measure their distance. His very
approximate estimations gave rise to a picture of the Universe
where the Milky Way is an ordinary stellar system of an
infinite number of other galaxies. In the 19th century, his son
John Herschel (1792-1871) continued searches for and
studies of `milky nebulae.' John Herschel expanded his
study to the southern hemisphere, doubled the number of
known faint nebulae, and continued studying their distribution 
in the sky.

The photographic process discovered in the middle of the
19th century allowed astronomers to abandon visual observations 
and to proceed to photographic sky surveys. At the end
of the 19th century, E.\,Barnard (1857-1923) started systematic 
photographic observations of the sky and performed the
first photographic survey of the Milky Way. Studies of the
structure and dynamics of our Galaxy greatly benefited from
the plan of `selected areas' by J.\,Kapteyn (1851-1922). To
execute this project, Kapteyn called upon astronomers from
across the world to carry out photographic observations and
to carefully study stars (including counting and measuring
apparent magnitudes, proper motions, etc.) in 206 areas evenly
distributed over the sky. This plan clearly demonstrated the
fruitfulness of international collaboration in solving laborious 
observational problems and anticipated some features of
big modern observational projects.

After E.\,Hubble (1889-1953) discovered the extragalactic
nature of faint nebulae, it became clear that the Universe is
much larger than had previously been thought. In order to
study the large-scale structure of the Universe and to understand 
the nature, origin, and evolution of its principal `bricks'
-- galaxies, extensive sets of extragalactic studies had to be
compiled and analyzed. This work was started by Hubble
himself (see, e.g., [3]), as well as by other astronomers
(Shapley, Ames, Humason, Lundmark, Bok, etc.).

\begin{figure}
\centerline{\psfig{file=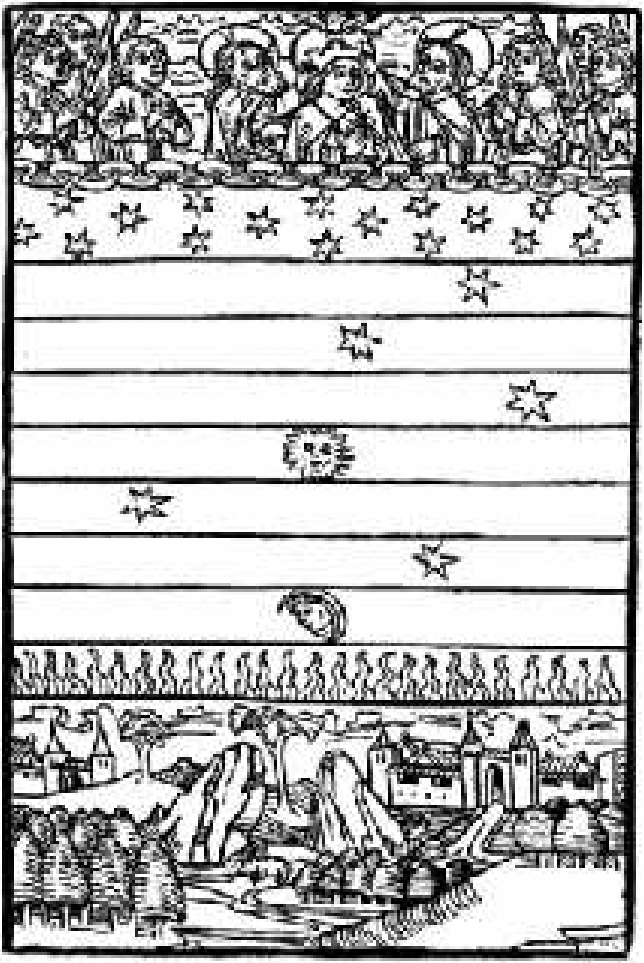,width=8.8cm,clip=}}
\caption{The structure of the Universe according to concepts of the 
middle of the 14th century [1]. At the bottom, the earth, water, air, 
and everything created from them are shown. The horizontal layers from 
bottom to top show: the sphere of fire, the sphere of seven planets' 
(the Moon, Mercury, Venus, the Sun, Mars, Jupiter, Saturn), the sphere 
of fixed stars, and divine heavens totally screened from observation 
by opaque clouds.}
\end{figure}

    At the end of the 1920s -- beginning of the 1930s, Hubble
performed a laborious survey of more than 40,000 galaxies
in 1,283 areas located in both celestial hemispheres [4]. The
main results of Hubble's work can be summarized as
follows: the number of galaxies continues to grow up to
the limiting magnitude of the survey; this growth is in
quantitative agreement with a homogeneous distribution of
galaxies in space (later, Hubble discovered that integral
counts of fainter galaxies increase more slowly than was
expected for their homogeneous distribution); there is a
strong dependence of the observed number of galaxies per
unit area of the sky on the galactic latitude (due to Galactic
extinction); it is the logarithm of the number of galaxies per
square degree ($N$) and not the number $N$ itself that, after
reducing all plates to standard conditions, follows a
Gaussian distribution. This last discovery was one of the
first indications that galaxies tend to `crowd' [5]. Hubble's
results strongly differed from star counts: in counting stars,
we reach the boundaries of our stellar system, while observations
of galaxies do not show the presence of a boundary of the
extragalactic world.

    During the entire subsequent history of the 20th century,
the main achievements in sky surveys and deep studies of
selected areas were due to `technological' successes, such a
the use of new big and specialized telescopes, increasingly sensitive
photo emulsions, and then CCD matrices and computers, the
elaboration of multi-object spectroscopy, etc. Each such a
`technological' step has led to ever deeper penetration into the
Universe.

    The invention of a new telescopic system, the wide-field
reflector, by Estonian optician Bernhard Schmidt (1879-1935) 
was one of the most important stages in the development of the 
observational technique. A correcting plate
mounted in front of a reflector's objective allows compensating 
for most aberrations of the main mirror. This opens the
possibility to construct candlepower telescopes with a wide
($\sim10^{\rm o}$) undistorted field of view. The best known Schmidt
telescope (the correction plate diameter is 122 cm, the
diameter of the objective is 183 cm) is installed at the Mount
Palomar Observatory in California. Its field of view is 6.$^{\rm o}$6. 
In 1950-1958, this telescope was used to carry out the famous
Palomar Sky Survey (see below). The biggest Schmidt
telescope (the correction plate diameter is 137 cm, the
diameter of the objective is 200 cm) is installed near Jena in
Germany. The biggest telescope of this kind on the territory
of the former Soviet Union operates at Byurakan Observatory (Armenia) 
(the correction plate diameter is 102 cm, the
diameter of the objective is 132 cm, the field of view is 4$^{\rm o}$). 
The Byurakan Schmidt telescope was used to carry out the well-known 
survey of galaxies with ultraviolet excess (the so-called
Markarian galaxies).

    In the last 10-15 years, several international projects
have been carried out that distinctively changed the aspect of
modern astronomy. Observational data on the structure of
our and other galaxies were increased by dozens and
hundreds of times. For the first time, it became possible to
study the evolution of galaxies and their large-scale structure
starting almost from the moment of their formation until
now. There are statements that a `golden age' of studies of
galaxy formation and evolution has begun. The general
feeling among astronomers and physicists (especially theoreticians) 
is partially characterized by the title of a colloquium
that took place at Caltech several years ago: ``Galaxy
formation: End of the Road!'' [6].

    Time will tell how justified such optimism is, but
undoubtedly extragalactic astronomy and observational
cosmology are in a period now that can possibly be
compared only with the 1920s, when the first galaxies were
identified and the expansion of the Universe was discovered.
In this paper, I attempt to briefly review selected observational 
projects of the last years. In view of the immense
observational data, I restrict myself to {\it optical} and 
{\it extragalactic} surveys.

    Throughout the review, I use a cosmological model with
$\Omega_m=0.3$, $\Omega_{\Lambda}=0.7$, and $H_0=70$ km/s/Mpc.

\section{General characteristics of surveys and deep fields}

Sky surveys and so-called `deep fields' represent different
strategies for studying extraterrestrial objects. There are no
strict criteria distinguishing deep studies of selected areas
and surveys. Guided by the characteristics of the most known
projects, the following empirical definitions can be proposed.

    {\it Sky surveys} include projects performing photometric 
and/or spectral observations of a significant fraction of the sky
(the total coverage $\geq$ 10$^4$ sq. deg.). The effective depth of 
surveys is $z\sim0.1$ (here and below, $z$ denotes redshift) or several
hundred megaparsecs (Mpc). Modern sky surveys are carried
out over several years by using, as a rule, middle-size
specialized telescopes.

    {\it Deep fields} relate to projects devoted to a detailed
exploration of relatively small sky areas (the characteristic
field coverage is $10^{-3} - 10^1$ sq. deg). Fields are much deeper
($z \geq 0.5$) compared to surveys and observations are performed 
with large telescopes. The typical exposures of a
deep field are $10^{-3} - 10^{-1}$ year.

`Integral' characteristics of some contemporary observational 
projects, many of which are discussed in detail below,
are plotted in Fig. 2 in the plane $B$ -- lg\,S (a) and 
$B$ -- lg\,(N/S) (b), where $B$ is the limiting magnitude of galaxies 
found in a survey or within a field in the 
$B$ filter\footnote{In some cases, the limiting magnitude value was 
approximately estimated from data in other color bands.}, 
S is the area on the sky (in square degrees), and N is the 
number of galaxies found.

    Figure 2 clearly illustrates the formal distinction introduced 
above between surveys and deep fields: the characteristics of modern 
projects are concentrated mostly in regions with S$\leq$1 sq. deg. and 
S$\geq$1000 sq. deg. This, of course, must be a temporary situation, and 
one can imagine a not-too-remote future when large fully robotic 
telescopes will measure galaxies with $B \approx 25^m - 30^m$ over 
most of the sky (the top right in Fig. 2a).

\begin{figure*}
\centerline{\psfig{file=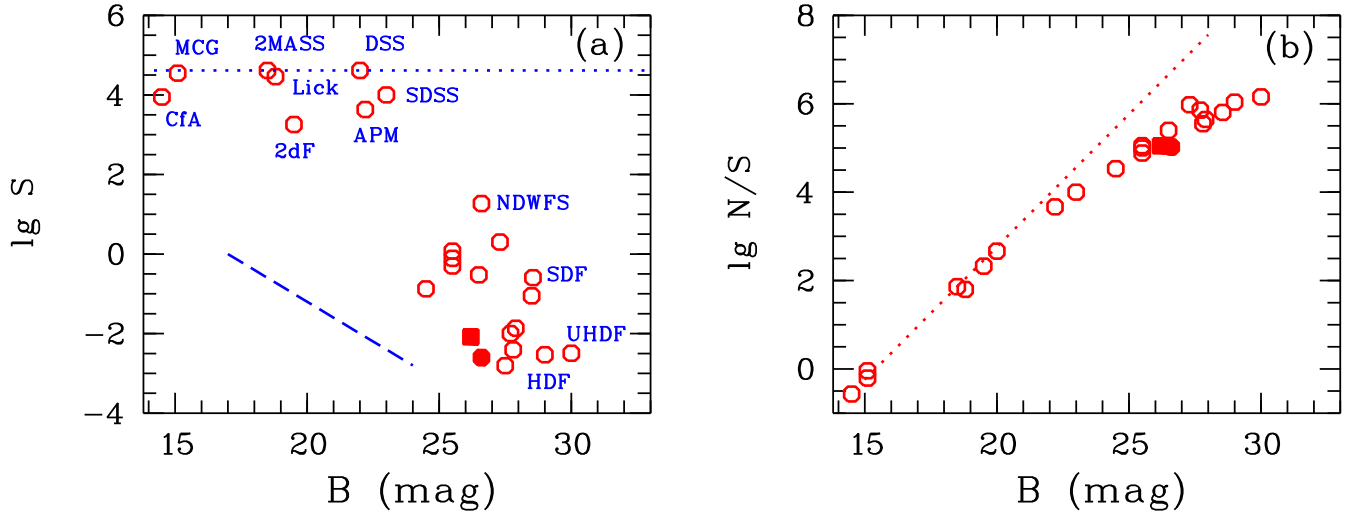,width=18cm,angle=-90,clip=}}
\caption{Characteristics of the main modern observational projects. 
The horizontal dotted line in Fig. (a) shows the total area of the sky. 
The black square and circle mark works performed with the 6-m SAO RAN 
telescope (see Section 4.9).}
\end{figure*}

    The dashed curve in Fig. 2a shows the simplest observational 
strategy with $E_{lim}$/S = const ($E_{lim}$ is the illumination
from the faintest objects detected). Such a dependence
between $E_{lim}$ and the area can be expected if observations
are carried out using one instrument with a fixed field of view
over a fixed total observation time. Big modern projects with
$B \geq 18^m$ show a steeper dependence in the $B$ -- lg\,S plane,
biased by the observations being made with larger telescopes
with a narrower, on average, field of view.

    Figure 2b shows the surface density of galaxies (the
number of galaxies per square degree) as a function of the
limiting apparent magnitude of the project. (It should be
borne in mind that the limiting magnitude values are
determined differently by different authors.) This figure
demonstrates that the number of galaxies per square unit
continuously increases up to $B \approx 30^m$. The old result obtained
by Hubble [7] is also clearly seen: the observed number of
galaxies increases with the limiting magnitude more slowly
than is expected for a homogeneous distribution in Euclidean
space (the dotted line in Fig. 2b). The reasons for such
behavior of galactic counts are the expansion of the Universe
and the evolution of galaxies with time.

     Figure 2b allows one to evaluate the number of galaxies
available for observation in the Universe. It can be seen that
the number of galaxies with $B \leq 30^m$ is about 1.5 $\times$ 10$^6$/sq. deg.
(i.e., one galaxy per each 3$''$ $\times$ 3$''$ square). Hence, the total
number of galaxies with $B \leq 30^m$ is $\sim10^{11}$.

     The actual volume of the Universe probed in a survey or a
deep field is determined, in addition to the deepness and area,
by a selection function -- a set of criteria used to select the
objects. The most widespread methods of object selection are
as follows [8]:

     (1) {\it Selection of all objects with a density flux above a fixed
threshold.} The detection limit is set, as a rule, in fractions of
the standard deviation of the night sky brightness fluctuations. The 
simplicity of this method allows a simple estimation of the space 
volume probed. The maximum redshift ($z_{max}$) at which an object with the 
proper (rest-frame) luminosity $L$ is still detectable at a given threshold 
$E_{lim}$ can be inferred from the relation 
$L = E_{lim} 4 \pi D_L^2$, where $D_L$ is the photometric distance depending 
on $z_{max}$. Then, the volume of a survey (deep field) is

$$
{\rm V}_{max}(z, L) = {\rm S} \int_{0}^{z_{max}} Q(z,\Omega_m,\Omega_{\Lambda}) d{\it z},
$$

where the function $Q$ depends on the cosmological model assumed
($\Omega_m$ and $\Omega_{\Lambda}$ are the relative contributions of matter
and vacuum energy to the total density of the Universe.)

    It should be noted that in practice, the selection is made
using not the observed flux densities but, to a greater extent,
the surface brightness of galaxies. For example, an object is
often believed to be detected if its flux in several neighboring
pixels in the CCD-image exceeds background fluctuations by
several times. Naturally, this procedure is biased in favor of
objects with a sufficiently high surface brightness.

    (2) {\it Selection by color indices.} This method accounts for
not only the observed flux but also the color indices, i.e., the
relative energy distribution in the galactic spectra. It is widely
applied to find the most distant galaxies, because their spectra
show a distinctive break near the Lyman limit (912 \AA) [9]. In
the past, this approach proved to be extremely effective in
discovering galaxies with ultraviolet excess (the Markarian
galaxies) and galaxies with active nuclei (see, e.g., [10]). The
calculation of the selection function and, correspondingly, the
space volume probed by observations using this method is
strongly dependent on the precise knowledge of the spectral
energy distribution in the objects under study.

    (3) {\it Selection by narrow-band observations.} The essence of
this method is the selection of galaxies that show an excess
when observed in narrow-band filters with respect to broad-band ones. 
This method is used to search for objects with
emission lines (star-forming galaxies, active galactic nuclei).
Observations are performed with narrow filters cutting
spectral ranges $\leq$100 \AA\,\, (to increase the contrast of the
emission object against the sky background) centered on the
wavelength (for example, L$\alpha$) corrected for the expected
redshift of a distant object. Clearly, in this case, the selection
function is determined by the equivalent width of the emission
line in the galaxy.

    A shortcoming of this approach is that galaxies are
searched for only in a very narrow interval of redshifts $z$ and
samples obtained in this way are relatively small. In addition,
only a small fraction of all galaxies from this redshift interval
is selected (namely, those that show a large equivalent width
of emission lines). These reservations restrict obtaining
statistically significant results on the general features of
distant galaxies.

    After the above comments, we turn to describing selected
projects. Projects similar to those described below are
currently being carried out at many observatories. Many
dozens of papers discussing the results of both new and old
surveys and deep fields are published each year. This diversity
of projects can be quite confusing (especially because many
projects have similar abbreviations). I therefore describe only
the principal projects playing an outstanding role in modern
astronomy.

    The main goal of the following `technical' description
(Sections 3 and 4) is to give the reader a flavor of the very
rapidly growing observational base of modern astronomy. A
distinctive feature of the last years is that the observational
data obtained are freely available for the scientific community
via the corresponding www pages.

\section{Sky surveys}

\subsection{Photographic surveys}

Photographic surveys performed with Schmidt telescopes
[11] had a great impact on the development of astronomy.
In the 1950s, a photographic survey of the sky available for
observations from California ($\delta > -33^{\rm o}$) was performed
using the 1.2-m telescope of the Palomar Observatory.
Almost a thousand plates 6.$^{\rm o}$5$\times$6.$^{\rm o}$5 each 
were obtained
in the blue and red spectral bands. Copies of the Palomar
sky survey (in the form of glass or printed copies of the
plates) were spread over most astronomical institutes in the
world and played a very important role in the development
of all fields of astronomy, from solar system studies to
remote galaxies and quasars [11]. Objects down to $B \sim 20^m$
can be distinguished in the Palomar prints, and the structure
of tens of thousands of galaxies with $B \leq 15^m$ can be
studied. In particular, based on the Palomar survey (its
official name is the Palomar Observatory Sky Survey I, or
POSS-I), catalogs of galaxies by Zwicky [12] and Vorontsov-Velyaminov 
(MCG in Fig. 2a) [13] were compiled. It is
by inspecting copies of this survey that systematic studies of
galaxies, from interacting [14] and double ones [15] to
galaxy clusters [12, 16], began.

     In the 1970s, the success of the Palomar survey stimulated
carrying out similar surveys of the southern sky by the 1.2-m
British Schmidt telescope (the Anglo--Australian Observatory (AAO), 
Australia) and the 1.0-m Schmidt telescope of
the European Southern Observatory in Chile. Due to great
progress in constructing telescopes and improving quality of
photographic emulsions, the limiting apparent magnitude of
these surveys (ESO/SERC) is by about 1.$^m$5 smaller than in
POSS-I. This, in turn, initiated, at the end of the 1980s, re-surveying 
the northern sky with the modified Palomar
Schmidt telescope using improved emulsions, but this time
with three filters, including the near infrared band centered on
$\lambda_{eff}\approx$ 8500 \AA. This survey was named POSS-II [17]. The
limiting magnitude in POSS-II for star-like objects is
$B \approx 22.^m5$.

     One photographic plate taken by a large Schmidt
telescope can have 10$^5$--10$^6$ images of stars and galaxies.
This restricted earlier works by visual inspection of only small
areas of the original plates. The effective reading of information 
from the Schmidt plates became possible only after high-speed measuring 
machines were designed that allowed image
digitizing and subsequent computer processing. It is in this
way that the first digital sky surveys APM and DSS appeared
at the beginning of the 1990s.

\subsection{APM}

In this project, the microdensitometer APM (Automatic
Plate Measuring machine) in Cambridge, England, was used
to scan 185 plates (the scan step was 0.$''$5) obtained with the
1.2-m Schmidt telescope of the Anglo--Australian Observatory (Australia) 
near the southern galactic pole [18]. The
plates cover $\sim$4300 sq. deg. on the sky. Around 20$\times$10$^6$ 
objects with $B \leq 22^m$ were discovered on these plates. For each
object, the coordinates, apparent magnitude, and a
dozen other parameters characterizing the brightness distribution and 
shape were determined. By analyzing photometric
brightness profiles, the objects were classified to form a
virtually complete sample of extragalactic objects containing
$\sim$2$\times$10$^6$ galaxies with $B \leq 20.^m5$.

\begin{figure*}
\centerline{\psfig{file=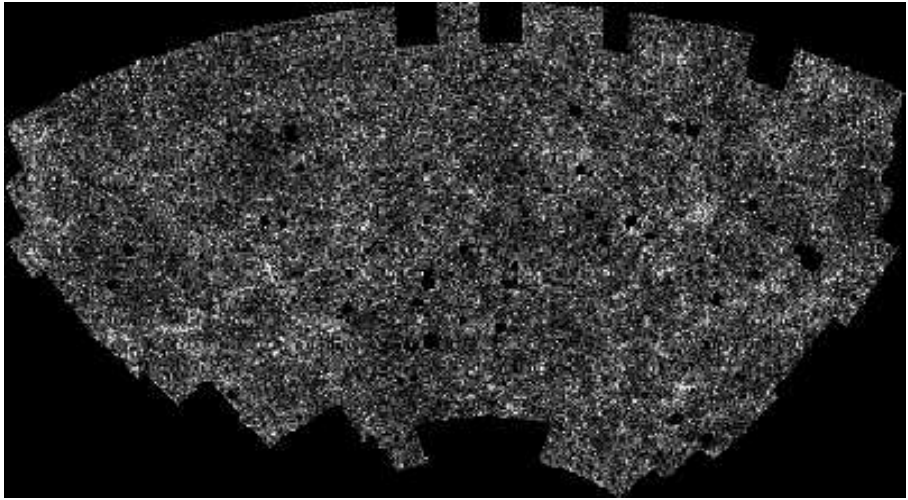,width=12cm,clip=}}
\caption{The visual distribution of APM galaxies.}
\end{figure*}

    The galaxy distribution of the APM survey over an area of
$\sim50^{\rm o}\times100^{\rm o}$ is shown in Fig. 3. Obviously, 
the projected
galaxy distribution is far from being homogeneous. Figure 3
clearly shows the presence of regions with an enhanced and
low concentration of galaxies and elongated structures. On
the basis of the APM survey, some important studies of the
large-scale distribution of galaxies were carried out (see, e.g.,
[19]) and the first objective catalog of galaxy clusters [20] was
compiled.

    The results of the APM survey and its extensions (in
particular, to the northern hemisphere) were used as the basis
for one of the most interesting projects of recent years -- the
2dF survey of galactic redshifts (see Section 3.5 below).

    In addition to APM, one can note several later projects
aimed at scanning the Palomar and ESO/SERC plates: for
example, APS (USA) [21] and COSMOS and SuperCOSMOS
(England) [22, 23]. These surveys are distinguished by a large
sky coverage and using plates in different colors.

\subsection{DSS and DPOSS}

DSS (Digitized Sky Survey) is the first high-quality and freely
available digitized image of {\it the entire sky} in the optical range.
This survey stemmed from the Space Telescope Institute
(STScI) project on creating a star catalog that can be used to
precisely point the Hubble Space Telescope (HST) to a
required object and guide it during observations [24]. To
compile such a catalog, the scanning of blue photographic
plates of the POSS-I and SERC surveys was initiated. The
scan step was 1.$''$7. Soon, it was understood that the
importance of the digitized images is far beyond the original
purpose and it was decided to open them to the wider
scientific community. The total amount of the original
version of the survey (DSS-I) reached 600 GB and, clearly,
such a huge amount of data was not easily transportable at
the beginning of the 1990s. However, after specially designed
tenfold compression of the data, the survey was fit into
120 CD-roms, which have been distributed through the
Astronomical Society of the Pacific (USA) starting in 1994
[25]. Later on, free access to DSS-I was open through the web
pages of STScI 
(http://archive.stsci.edu/dss/). Now, this
survey can be remotely accessed through several cites in the
USA and Canada, as well as in some European countries and
Japan.

      The DSS-II survey was the natural extension of DSS-I
using data from POSS-II [26]. The POSS-II plates of the
northern sky, as well as the SERC plates and other surveys of
the southern sky, were scanned with the step 1.$''$0. Plates in
three color bands were digitized, which allowed a comparison
of sky areas in different spectral bands. The total volume of
DSS-II attains $\sim$5 TB and remote access to it is available, as
a rule, through the same web pages as for DSS-I.

    From the very beginning, DSS became one of the most
useful and required tools of modern astronomy. It allowed
obtaining an image of any area of the sky in several seconds,
which strongly facilitated the preparation and planning of
observations. Using the DSS images, a great number of
papers devoted to study of individual galaxies, galaxy
groups, their large-scale distribution and geometrical 
characteristics, the optical identification of objects observed at
other wavelengths, etc. have been published.

    The DPOSS project (Digitized Palomar Observatory Sky
Survey -- http://dposs.caltech.edu) is also based on scanning
the POSS-II plates in three colors; however, the subsequent
data processing and calibration are different from those used
in DSS [27]. This survey covers all the northern sky with
$\delta > -3^{\rm o}$. 
Extensive CCD-observations were performed at the
Palomar Observatory to provide photometric calibrations of
this survey. DPOSS includes a database of images scanned
with the step 1.$''$0 ($\sim$3 TB) and several catalogs based on
these data. The ultimate goal of DPOSS is the creation of the
PNSC catalog (Palomar Norris Sky Catalog) including all
objects found in the survey up to the limiting magnitude
$B \approx 22^m$. More than a hundred measured parameters will be
provided for each source in the catalog, and objects with
$B \leq 21^m$ will be classified as stars or galaxies. It is expected
that PNSC will provide information on $> 50 \times 10^6$ galaxies
(including $\sim10^5$ quasars) and $> 10^9$ stars.

    The surveys mentioned above were based on photographic 
observations and, naturally, suffer from all the
standard shortcomings of photo emulsions, such as low
sensitivity, restricted dynamical range, and nonlinearity. All
projects (both surveys and deep fields) that we discuss below
are truly digital, because CCD detectors are used to perform
them.

\subsection{2MASS}

\begin{figure*}
\caption{The distribution of the 2MASS survey objects in the sky (in
galactic coordinates).}
\end{figure*}

One of the best known digital surveys in a wavelength range
close to the optical one is 2MASS (Two Micron All Sky
Survey), which is the result of the collaborative efforts of the
University of Massachusetts and the Infrared Processing and
Analysis Center at Caltech (http://www.ipac.caltech.edu/2mass). 
2MASS is a purely photometric survey covering the
whole sky in filters $J$ (1.25 $\mu$m), $H$ (1.65 $\mu$m), and 
$K_s$ (2.17 $\mu$m) [28]. Observations were carried out from June
1997 to February 2001 with two robotic 1.3-m telescopes in
Arizona, USA, and Chile. Each instrument was equipped
with a three-channel camera imaging the sky simultaneously
in the three spectral bands using 256$\times$256 IR CCD-detectors
with the pixel size 2.$''$0.

    The survey consists of 10 Tbt of images and catalogs of
objects. Calibrated images of any area of the sky in the $J$, $H$,
and $K_s$-bands are available through several sites (for example,
http://irsa.ipac.caltech.edu/). An atlas of near-IR images of
864 galaxies has been published [29]. The point source catalog
(PSC) lists coordinates and photometric data for about
500 million objects (mostly Milky Way stars). The extended
source catalog (XSC) includes data for $\sim$1.65 million objects
with angular sizes $\geq$10$''$--15$''$ [30]. More than 98\% of these
objects are galaxies, others are HII regions, stellar clusters,
planetary nebulae, etc. The limiting magnitudes of the
extended objects from the XSC are 13.$^m$5 (2.9 mJy) and
15.$^m$0 (1.6 mJy) in the $K_s$ and $J$-bands, respectively. Figure 4
shows the distribution of objects from the XSC in galactic
coordinates. The letters in this figure mark clusters and
superclusters of galaxies, the extended image in the center is
the Milky Way stellar disk seen `edge-on' [31].

    Over the few years since completion, the 2MASS survey
has already greatly impacted the development of astronomy.
For example, according to the Astrophysics Data System
(ADS: http://adsabs.harvard.edu) [32], the number of published 
papers that used the 2MASS data approached a
thousand by the beginning of 2005. The main areas of study
benefiting most from using the 2MASS data include the large-scale 
structure of the Milky Way and distribution of galaxies
in the nearby Universe (in the optical range, such studies are
strongly restricted by galactic interstellar absorption), as well
as searches for and explorations of new types of astronomical
objects (for example, low-mass stars, brown dwarfs, `red'
quasars, etc.).

\subsection{2dF and 6dF}

The 2dF (2 degree Field Galaxy Redshift Survey, or
2dFGRS) represents a spectroscopic survey of $\sim$5\%
($\sim$2000 sq. deg.) of the sky performed by British and Australian
astronomers with the 3.9-m telescope of AAO [33, 34]. (The
coverage of this survey is rather small to classify it as a survey,
according to Section 2, and clearly demonstrates the conditionality 
of dividing modern projects into surveys and deep
fields.) Objects for this survey were sampled using the
extended APM source catalog (see Section 3.2) and included
galaxies brighter than $B \approx 19.^m5$ near the North and South
galactic poles. A specially designed multi-object spectrograph
allowing simultaneously obtaining spectra of 400 objects
within the 2$^{\rm o}$ field of view was used. Observations included
272 nights in the period between 1997 and 2002.

      The openly accessible results of the project 
(http://www.mso.anu.edu.au/2dFGRS/) include: a photometric
catalog of objects selected for spectroscopic studies; a
spectroscopic catalog of 245,591 objects listing redshifts z
and spectral types (221,414 galaxies in this catalog have
reliable redshift measurements); and special software to
fetch both fits-files with spectra and subsamples of objects
according to selected criteria.

      Upon its completion, the 2dF survey became the largest
galaxy redshift survey, with the number of measurements
exceeding 10$^5$ for the first time. It allows investigating the
three-dimensional large-scale structure of the surrounding
Universe with an unprecedented accuracy. The median redshift 
of the survey is $z=0.11$, which corresponds to the
photometric distance $\sim$500 Mpc. As an example, in Fig. 5,
we show the distribution of 63,000 galaxies from the survey
located within narrow 3$^{\rm o}$-bands around the north (to the left)
and south (to the right) galactic poles [35]. In contrast to Figs 3
and 4, which show the galaxy distribution projected on the
sky, this figure plots the galaxy distribution along the line of
sight. The decrease in the number of galaxies with increasing $z$
is the result of sampling by apparent magnitude biasing only
bright remote galaxies. The principal elements of the 
large-scale structure of the Universe -- clusters and superclusters,
filaments, and voids -- are clearly seen in Fig. 5.

\begin{figure*}
\centerline{\psfig{file=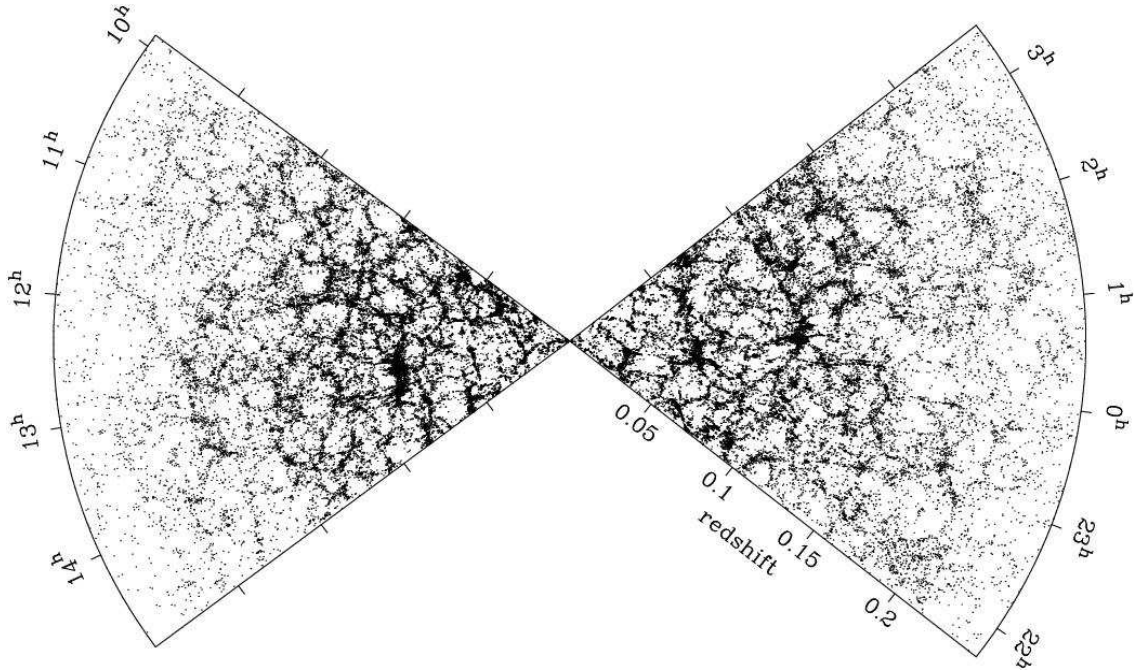,width=15cm,clip=}}
\caption{The distribution of galaxies in the 2dF survey field.}
\end{figure*}

    The 2dF survey plays a major role in modern astronomy
as one of the main sources of information on the spatial
distribution of galaxies and the density of matter in the
Universe. For example, the data from the survey allowed the
upper limit of the total mass of the neutrino 
$m_{\nu,tot} < 1.8$ eV to
be derived [36] and, in combination with data on the cosmic
microwave background, the values of the main cosmological
parameters to be improved [37].

    An autonomous subproject within the 2dF is the 2dF QSO
Redshift Survey (2QZ), a survey of redshifts of quasars
located near the north and south galactic poles (the total
coverage is $\sim$700 sq. deg.) [38]. The 2QZ contains spectra of
23,338 quasars, 12,292 stars of our Galaxy (including around
2,000 white dwarfs), and 4,558 emission-line galaxies. The
data, including photometry, redshifts, and fits-files with
spectra, are available via http://www.2dfquasar.org/.

    The 6dF (6dF Galaxy Survey, or 6dFGS) represents a
survey of redshifts and peculiar velocities of galaxies selected
mainly from the XSC 2MASS survey catalog (see Section 3.4)
[39]. Selecting galaxies in the IR spectral range, where the
effect of interstellar absorption inside the Milky Way is much
smaller than in the optical range, allows much better studies
of the object distribution at low galactic latitudes. The 1.2-m
Schmidt telescope of the Anglo--Australian Observatory is
used, equipped with a multi-object spectrograph simultaneously 
taking spectra of 150 objects inside the telescope's
6-degree field of view. Redshifts of around 150,000 galaxies
are planned to be measured. The survey will cover almost the
entire southern sky with 
$\delta < 0^{\rm o}$ (the survey coverage is
$\approx$17000 sq. deg.) and will give detailed information on the
distribution of galaxies within the nearby ($z \approx 0.05$) volume
of the Universe. At the beginning of 2005, about half of
the input galaxy sample was made available 
(see http://www-wfau.roe.ac.uk/6dFGS/.)

    The principal goal of the 6dF is to study large-scale
deviations in the velocity of galaxies from the homogeneous
Hubble expansion. The distribution of such deviations
provides the unique means to study mass distribution in the
Universe independent of the assumptions that galaxies follow
the true mass distribution. For about 15,000 early-type
galaxies evenly distributed over the southern sky, $z$-independent 
distances will be determined using the Fundamental
Plane method (the Fundamental Plane is a three-parameter
relation between photometric and kinematic characteristics
of galaxies, see, e.g., [40]). Then, by comparing these distances
with those derived from the observed values of $z$, it will be
possible to estimate the peculiar velocities of galaxies arising
due to inhomogeneities in mass distribution. (In this way, the
Great Attractor with the mass         
$\sim 5 \times 10^{16}$\,M$_{\odot}$
in a relatively nearby region of the Universe was found [41]).

\subsection{SDSS}

The Sloan Digital Sky Survey (http://www.sdss.org/) is often
referred to as one of the most grandiose astronomical projects
in history. Starting at the end of the 1980s, it is being carried
out by more than a hundred scientists from the USA, Japan,
and some European countries [42, 43].

    The purpose of the SDSS is to perform a photometric and
spectral study of a quarter ($\approx$10000 sq. deg.) 
of the sky. The
survey covers one large area near the Northern Galactic Pole
and three bands (with a total coverage of 740 sq. deg.) in the
southern hemisphere. Observations are carried out with a
specially designed 2.5-m telescope (the modified 
Ritchey--Chretien system, 3$^{\rm o}$ field of view) in 
New Mexico (USA). The
telescope is equipped with a CCD-camera and a couple of
identical multi-object fiber-optic spectrographs to simultaneously 
take spectra of 640 objects. Auxiliary works are also
performed with several other telescopes.

   The main goal of the photometric observations is to
construct a database for $\sim$10$^8$ galaxies and $\sim$10$^8$ stars,
containing the precise ($\leq0.''1$) coordinates and photometric 
and other characteristics. The observations are
carried out in five broadband filters in the spectral range
from 3500 \AA\,\, to 9100 \AA. The limiting magnitude of the
photometric survey is $B \approx 23^m$ for point-like objects.

     The spectral observations should provide spectra of about
$10^6$ galaxies, $10^5$ quasars, and $10^5$ stars selected from the
photometric survey. Two samples of galaxies have been
selected: $\sim$900,000 galaxies with $B\leq19^m$ and $\sim$100,000
objects with large color indices (`red' galaxies) and
$B \leq20.^m5$. The quasar candidates are selected using the
observed color indices. Stellar-like objects with radio emission 
are also included in the quasar candidate sample.

    Systematic observations for this project started in April
2000 and are to be completed by summer 2005. In the course
of the observations, the processed data have become available
via the web pages of the survey. By the beginning of 2005,
around half of the survey was made available. These data
include $\sim$6 TB of images, the photometric and astrometric
catalogs for 1.41$\times$10$^8$ objects, and the spectra of 528,640
objects, including 374,767 galaxies, 51,027 quasars, and
71,174 stars [44]. The final results of the SDSS will be
presented in 2006.

    The SDSS is not yet completed, but a great number of
papers (more than a thousand, according to ADS) have
already been published. These studies include all fields of
optical astronomy, from asteroids (more than several dozen
thousand of them have been already discovered by the SDSS)
to quasars (quasar redshifts measured in the survey first
exceeded the $z=6$ barrier) [45]. The SDSS and 2dF data
increased by hundreds of times the observational information
on the structure, spectral characteristics, and spatial 
distribution of galaxies in the nearby ($z \leq 0.2$) volume 
of the Universe.
At present, almost all characteristics of galaxies known earlier
are being revised and improved. This relates in particular to
their spatial distribution, the luminosity function, and the
dependence of galactic properties and the star formation rate
on the environment (see Section 5 below).

\section{Deep fields}

This section describes some remarkable projects to study
relatively small ultra-deep areas -- the deep fields (see the
lower right corner in Fig. 2a). I will restrict myself to works
carried out in the last ten years. Earlier papers are
referenced in reviews by Koo and Kron [46], Sandage [47],
and Ellis [48].

\subsection{WHDF}

The WHDF (William Herschel Deep Field) is the deepest
ground-based image of a small area taken with a middle-class
telescope. The $7' \times 7'$  patch of sky centered on 
$\alpha(2000)=00^h22^m33^s$ and $\delta(2000)=+00^{\rm o}21'$
was imaged in
1994--1997 with the 4.2-m William Herschel telescope (the
Roque de Los Muchachos Observatory, Spain) [49]. Several
dozen thousand CCD-frames were obtained in each of four
filters ($U$, $B$, $R$, and $I$) with the respective total exposure time
34, 28, 8, and 5 h. A careful processing and summing-up of
frames enabled reaching the $B\approx28^m$ limiting magnitude 
for unresolved objects. The number of galaxies discovered in the 
WHDF is around six thousand. This field was later
imaged in the IR bands by other telescopes.

     The results of the WHDF studies have been used to
analyze the dependence of the number of galaxies on their
apparent magnitude (this provides information on the
early cosmological evolution of galaxies), to determine the
surface (angular) correlation function of galaxies, and to
select the most distant objects by using color indices (see
Section 2) [49, 50].

\subsection{HDF}

The deep fields of the HST [their standard names are the
Hubble Deep Field North (HDF-N) and Hubble Deep
Field South (HDF-S)] are likely to be the most well-studied
patches of the sky. Their exploration has led to some
important discoveries on the structure and evolution of
distant galaxies [51].

    Observations with the HST (the diameter of the main
mirror is 2.4 m, the Ritchey ± Chretien system) in the first half
of the 1990s demonstrated that this instrument resolves the
structure of distant galaxies and these galaxies look different
than those at $z \approx 0$. The idea emerged to use some free time at
the discretion of the STScI director (at that time, Robert
Williams) to obtain an unprecedented deep image of one
typical area at high galactic latitudes [52]. The area was very
carefully selected: the interstellar absorption in the Milky
Way in its direction had to be small, the area should not have
had very bright (in all spectral bands) objects, no relatively
close galaxy clusters should have been present, etc. As a
result, a region of the sky in Ursa Major was selected.

\begin{figure*}
\centerline{\psfig{file=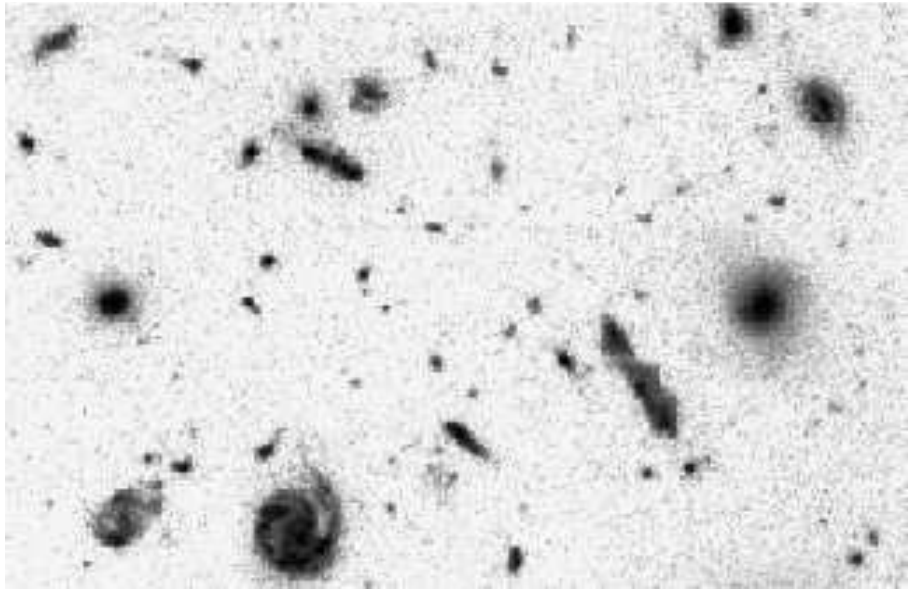,width=12cm,clip=}}
\caption{A fragment ($\sim 20'' \times 30''$) of the HDF-N.}
\end{figure*}

    Observations of this northern area (HDF-N) were carried
out in December 1995 with the WFPC-2 (Wide-Field
Planetary Camera 2) in four broadband filters centered on
3000 \AA\, (filter F300W), 4500 \AA\, (F450W), 6060 \AA\, (F606W),
and 8140 \AA\, (F814W). From fifty to a hundred individual
frames were taken, with each frame being taken at a slightly
displaced position of the telescope (such that one object fell
onto different elements of the CCD matrix). Such a dithering
technique allowed obviating the detector's defects and
constructing integral deep field images with the step 0.$''$04,
which is smaller than the matrix pixel size. The HDF-N covers
about 5.3 sq. min., and the field has a nonrectangular shape. The
total exposure time in each filter amounted to one to almost
two days. The HDF-N was later observed in the near IR
bands.

    In mid-January of 1996, immediately after primary data
processing, the deep field images were open to access through
the STScI web-pages 
(http://www.stsci.edu/ftp/science/hdf/hdf.html). 
In addition to original images, this site provides a
detailed description of observations, their processing, and
calibration. It turned out that in the HDF-N, one can detect
galaxies as faint as $B\sim29^m$. Depending on the selection
criteria, up to 2000--3000 galaxies in this field can be
discovered (for comparison, only several dozen stars of our
Galaxy were found). The main feature of the HST deep field
is, of course, a much better resolution ($\approx$0.$''$1) than can be
achieved by ground-based observations. Such an angular
resolution enables one to study the observed structure of
distant ($z \geq 1$) galaxies with the linear resolution $\leq$1 kpc.
The principal shortcoming of the field is its small angular size
(at $z=1$, it corresponds to only $\approx$2 Mpc in the comoving
frame), and therefore the statistics of objects in the field are
not representative.

    The first reproductions of the HDF-N (see, e.g., Fig. 6)
clearly demonstrated that the Universe looked differently
several billion years ago. Distant galaxies are much more
asymmetric, and there are numerous interacting and irregular
systems. The Hubble field provided very rich data enabling
the exploration of the morphology and sizes of galaxies, deep
galaxy counts, searches for extremely distant objects
($z\geq5$), etc.

    The obvious success of the HDF-N stimulated carrying
out an analogous project in the southern hemisphere [53].
Observations of the southern area (located in the Tucanus
constellation) were carried out in October 1998. The HDF-S
project has two important differences from the HDF-N: a
remote quasar with $z=2.24$ falls within this field and several
instruments are simultaneously used. The following results
were obtained: (1) a deep image of the field was obtained with
the WFPC-2 (with the same filters and using the same
methods as in the HDF-N), (2) the spectroscopy of the
quasar in the spectral range 1150 to 3560 \AA\, was made, an
ultradeep image of a small area near the quasar was taken
with the STIS (Space Telescope Imaging Spectrograph), and
(3) a small area was imaged with the NICMOS (Near Infrared
Camera and Multi-Object Spectrometer) at 1.1 $\mu$m, 1.6 $\mu$m,
and 2.2 $\mu$m. All three instruments have different fields of
view, and therefore the instrumental field must be specified in
referencing the HDF-S. As in the case of the HDF-N,
completely reduced observations of this field were open to
access at the end of November 1998 
(see http://www.stsci.edu/ftp/science/hdfs.html).

    The policy of providing open access to the data, as well as
their uniqueness (both fields had been the deepest `punctures'
into the Universe in several years), ensured that they were
enormously popular and in demand. Both HDF areas were
observed many times from the ground and from space and in
all other spectral ranges, from X-ray to radio. According to
ADS, at the beginning of 2005, the data of these fields were
used in at least fifteen hundred papers.

\subsection{CDF}

\begin{figure*}
\centerline{\psfig{file=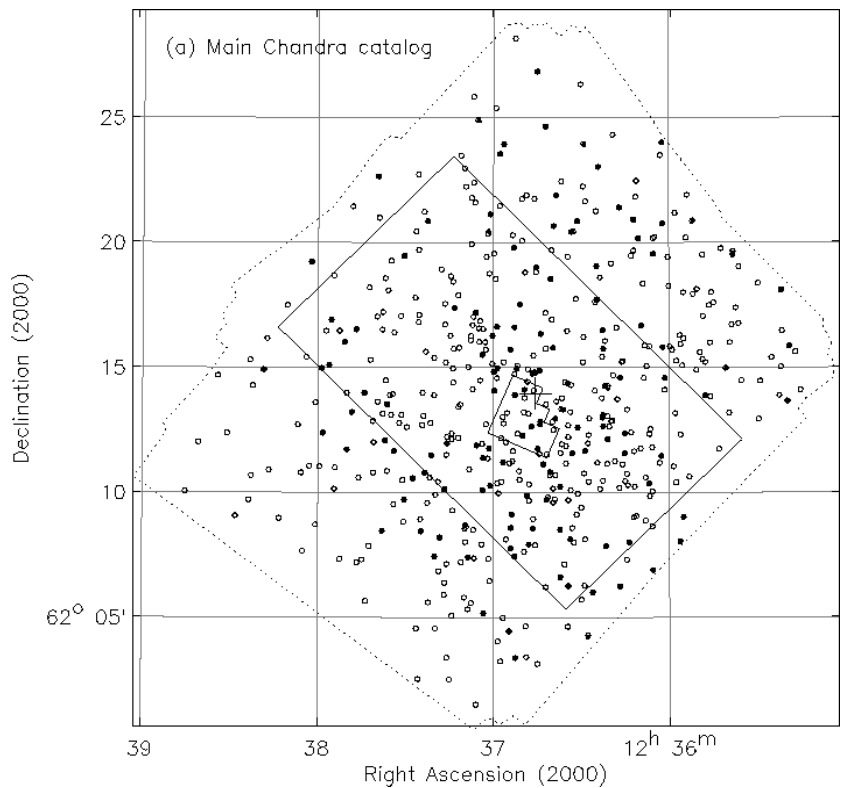,width=11cm,clip=}}
\caption{The distribution of X-ray sources within the CDF-N (the
boundaries of this field are shown by the dotted line). The white circles
show sources detected with a 10$^6$ s exposure, the black circles stand for
fainter sources that appeared when increasing the exposure time to
2$\times$10$^6$ s. The small polygon in the center marks the HDF-N field.}
\end{figure*}

The Chandra Deep Fields represent a series of deep (exposure
$\sim$10--20 days) images of small patches of the sky obtained by
the space Chandra X-ray Observatory (CXO). The best
known of them are the North field (CDF-N) and South field
(CDF-S). Observations of the CDF-S in the constellation
Fornax were carried out in 1999--2000. The total exposure
time was around 10$^6$ s. The CDF-S covered 0.109 sq. deg. with the
pixel size 1$''$. In the Southern Chandra field, 346 sources were
discovered in the energy range 0.5--7 keV [54, 55]. Most of
these sources are extragalactic objects, mainly active galactic
nuclei and star-forming galaxies.

    The CDF-N was the deepest X-ray image of the sky area
obtained by the beginning of 2005. The field covers 0.124 sq. deg.
with the total exposure time approaching 2$\cdot$10$^6$ s. The
coordinates of the Northern Chandra field are close to those
of the HDF-N (see Section 2.4), but the area of the CDF-N is
many times larger than HDF-N (Fig. 7). In the CDF-N,
about 600 X-ray sources in the energy range 0.5--8 keV were
discovered [56, 57]. Twenty of these sources lie within the
HDF-N (Fig. 7).

    Spectroscopic studies of optical counterparts of the
CDF-S and CDF-N X-ray sources revealed that they lie at
$z \sim 1$, with some sources having $z > 3$ (see, e.g., [57]). 
For the first time, the CDF data enabled evaluating the evolution of
the X-ray luminosity of active galactic nuclei as a function of $z$
and investigating X-ray properties of normal galaxies at large
redshifts [58].

\subsection{FDF}

The FORS Deep Field is devoted to a detailed photometric
and spectroscopic study of a $\sim 7' \times 7'$ area located in the
vicinity of the south galactic pole with the FOcal Reducer/low
dispersion Spectrograph (FORS) on the 8.2-m ESO VLT
telescope [59]. The main photometric observations were
carried out in 1999--2000 with the UT1 (Antu) ESO VLT
telescope with five broadband filters spanning the spectral
range from $\sim$3700 to $\sim$8000 \AA. With each filter, several
dozen images were taken, and more than a hundred were
obtained with the $R$ filter. The total exposure time varied
from 6 to 12 h depending on the color band. The image
quality in the integral image is better than 1$''$ in each filter.
Near-IR observations ($J$ and $K$ filters) of the field were also 
carried out by the ESO NTT telescope.

     A careful analysis of the FDF images allowed discovering
almost 10,000 objects, mostly galaxies (only 50 stars are seen
in the field). The limiting magnitude in the FORS field with
filter $B$ turns out to be comparable with that in the HDF (see
Section 4.2), and is smaller than the HDF only by $\sim$1$^m$ with
the other filters. The ESO NTT telescope has started a
spectral survey of the FDF objects. By the present time,
redshifts for several hundred galaxies from this field have
been reported (see e.g., [60]).

     Additional observations of the FDF and data analysis
continue. Some preliminary results on the physical properties
of distant ($z > 3$) galaxies, their luminosity function 
evolution, the history of star formation in the Universe, etc., are
reported in [61].

\subsection{SDF and SXDS}

The main goal of the Japanese project Subaru Deep Field
(SDF) is to select and study a large sample of distant ($z > 4$)
galaxies. Since 1999, the Japanese Subaru (Pleiades) telescope
has been carrying out multicolor photometric and spectral
observations of a $34' \times 27'$ area near the north galactic pole.
Photometric data were obtained in five broadband 
($B$, $V$, $R$, $i'$) and two narrowband filters in the near-IR 
wavelength range centered on 8150 \AA\, and 9196 \AA\, [62]. The 
central $2' \times 2'$ part of the SDF was also observed in $J$ 
and $K$ bands [63]. The total exposure time in each band was
approximately 10 h with the limiting magnitudes near
28.$^m$5 in the $B$ filter and 23.$^m$5 in $K$. More than 150,000
objects (mostly galaxies) were discovered in the Subaru Deep
Field. The processed SDF images and catalogs of objects are
available through http://soaps.naoj.org/sdf/.

\begin{figure}
\centerline{\psfig{file=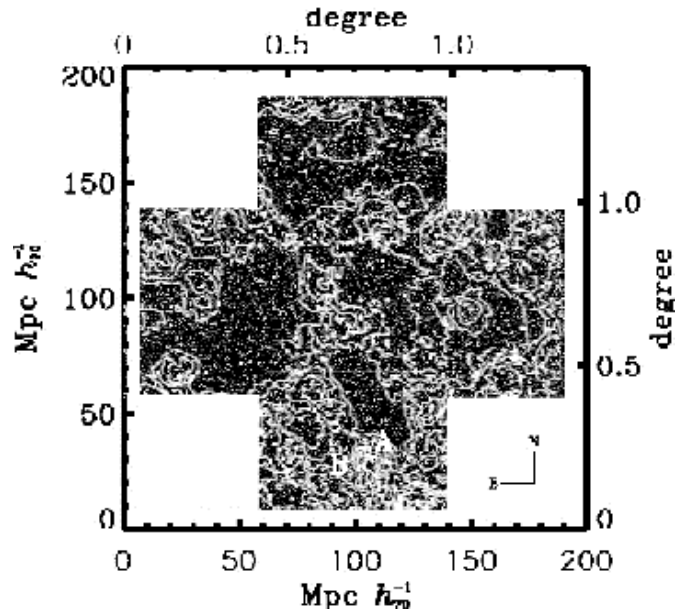,width=8.8cm,clip=}}
\caption{The distribution of L$\alpha$-emitting objects with 
$z=5.7 \pm 0.05$ in the SXDS (the light contour lines).}
\end{figure}

    Many interesting results on the properties and distribution 
of distant galaxies were obtained using the SDF. In
particular, several objects with $z \geq 6$ were discovered (their
age is only several million years), including the most distant,
at the present time, spectroscopically confirmed 
L$\alpha$-emitting object with $z=6.60$ [64].

    The Subaru/XMM-Newton Deep Survey (SXDS) is a
multi-wavelength survey of a small ($\sim 1.3$ sq. deg.) area 
obtained with several ground-based and space instruments (Subaru,
UKIRT, XMM-Newton, VLA, GALEX, JCMT) (see,
e.g., [65]). The SXDS is centered at 
$\alpha(2000)=2^h18^m00^s$ É $\delta(2000)=-5^{\rm o}00'00''$.
This survey is less deep (in terms
of the limiting magnitude) than the SDF, but has a
coverage several times larger and, in addition, allows 
multi-wavelength studies of objects.
    
At the end of 2003, photometric observations of the
SXDS field were performed in a narrow band centered on
$\lambda = 8150$\,\AA. Using a special method based on comparison of
the energy flux density in different spectral intervals, more
than 500 candidates in galaxies with $z = 5.7 \pm 0.05$ were
selected [66]. Spectral observations confirmed that most
objects must actually reside at such $z$. From analysis of the
space distribution of galaxies, the authors of [66] arrived at
the conclusion that some primordial large-scale structure of
galaxies with extended filaments, voids, and even with two
clusters under formation (see Fig. 8) must already exist at
$z=5.7$ (!).

\subsection{COMBO-17}

COMBO-17 (Classifying Objects by Medium-Band Observations in 
17 filters) represents a multicolor photometric survey
of five $\sim0.^{\rm o}5\times0.^{\rm o}5$
areas (including the CDF-S, the south
galactic pole, and the Abell\,901/902 supercluster) obtained
with the 2.2-m MPG/ESO telescope in Chile [67,68]. The
main feature of this project is that observations were carried
out with 17 filters (five broadband -- $U$, $B$, $V$, $R$, $I$ , 
and 12 medium-band) covering the spectral range 3500--9300 \AA.
Such a detailed photometry allows constructing a kind of
low-resolution spectrum of each object, which can be used
both to spectrally classify and to evaluate the redshift of each
object with a relatively good accuracy ($\sigma_z \approx 0.03$). 
The relatively small depth of COMBO-17 ($B \approx 25.5$ [68]), 
therefore, is compensated by the large coverage, as well as by the
possibility of estimating the type and $z$ of an object without
additional spectral observations. These advantages of
COMBO-17 make it a convenient tool to study galaxies
using the weak gravitational lensing method [69].

    The results obtained in this project for the Southern
Chandra Field, including a catalog of 63,501 objects, were
recently published in [70] 
(see also http://www.mpia.de/COMBO/combo$\_$index.html).

\subsection{GOODS}

In January 2004, a special issue of the {\it Astrophysical Journal
Letters} was entirely devoted to preliminary results obtained in
the GOODS (Great Observatories Origins Deep Survey)
project [71]. The GOODS is a new-generation project after
the HDF, combining deep multi-wavelength observations
from several space (HDF, SIRTF, CXO, XMM-Newton)
and ground-based (ESO VLT, ESO NTT, KPNO 4-m, etc.)
telescopes. The scientific goals of the GOODS include
estimations of stellar and dynamical masses of bright
galaxies up to $z \sim 5$, measurements of the star formation
rate in complete samples at different $z$, studies of the origin of
the Hubble sequence, measurements of the relative contributions 
of stars and active galactic nuclei to the global energy
budget of the Universe, and studies of individual sources
contributing to intergalactic background radiation in all
spectral ranges.

    Observations have been carried out in two areas
$\sim$160 sq. min., each almost centered on the HDF-N (Section
4.2) and CDF-S (Section 4.3).

    The fields were observed with the HST by the Advanced
Camera for Surveys (ACS) installed in 2002 in four broad-band 
filters F435W ($B$), F606W ($V$), F775W ($i$), and F850LP ($z$)
(the three figures in the filter's name indicate the central
wavelength in nm). As with the HDFs, dozens of dithered
exposures are taken with each filter. Observations in $V$, $i$, and
$z$ were carried out during five periods delayed by 40--50 days.
(Such an observational strategy was adopted to facilitate
searches for distant `cosmological' supernovae. As a result,
more than 40 supernovae were discovered by the GOODS,
with six SN Ia's at $z > 1.25$ [72].)

    After initial reductions and the superposition of individual 
frames, integral images of the fields in all filters were
obtained with the step 0.$''$03 (the actual angular resolution in
the images is $\approx0.''1$). The limiting magnitude of
extended objects in these fields is by 0.$^m$5--0.$^m$8 worse
than in the previous HST deep fields, but the total coverage
of the GOODS is 30 times larger than that of the HDF-N
and HDF-S taken together. The original HST frames and
reduced images are available through the web pages of the
GOODS project: http://www.stsci.edu/science/goods/.

    Observations in the framework of the GOODS project
with other telescopes (including, in particular, Keck and
Gemini) either have already been made or are being carried
out. See paper [71] or the project web page for details.

\subsection{HUDF}

\begin{figure}
\centerline{\psfig{file=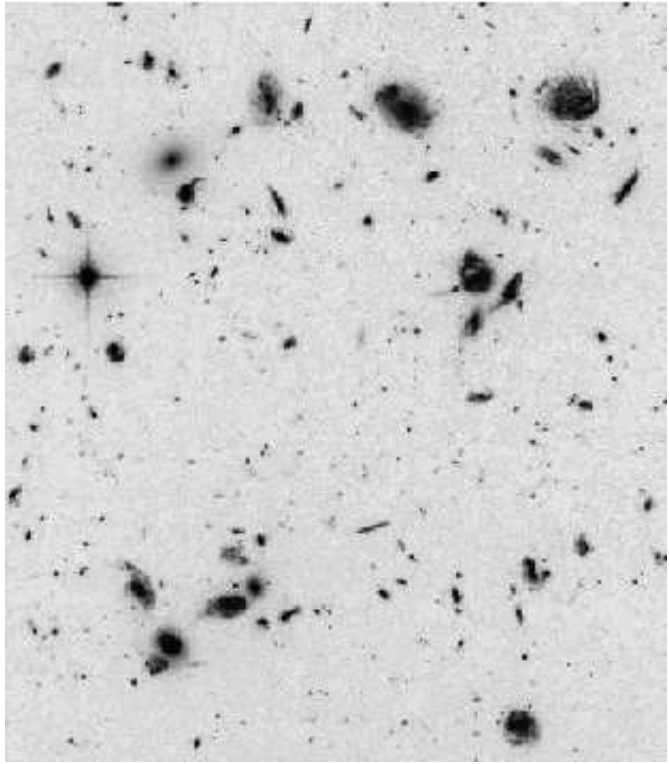,width=8.8cm,clip=}}
\caption{ Detail $\sim 50'' \times 55''$ of the HUDF.}
\end{figure}

The Hubble Ultra Deep Field (HUDF) is the deepest optical
imaging of a patch of the sky ever made (Fig. 9). The authors
of the project believe it will remain such in the next several
years and, consequently, this field will long remain the main
source of information on the most distant objects in the
Universe [73].

    The HUDF is located within the limits of the Southern
Chandra Field (CDF-S) and, hence, within the GOODS field.
The precise coordinates of the HUDF are 
$\alpha(2000)=3^h32^m39^s.0$ and $\delta(2000)=-27^{\rm o}47'29.''1$.
Major observations were carried out by the HST from September 2003 till
January 2004 with a wide-field camera (WFC) ACS with the
same four filters as the GOODS observations. The field
coverage is relatively small: 11.5 sq. min. More than a hundred
individual images were taken in the $B$ and $V$ filters
with the total exposure time $\sim$40$^h$.
Observations in the $i$ and $z$ bands include almost 300
frames with the total exposure time $\sim$100$^h$ with each filter.

    The final calibrated HUDF images with the step 0.$''$03
(the image size with each filter is 430 Mb) and the catalog of
discovered objects can be found on the web page of the
project: http://www.stsci.edu/hst/udf. The HUDF is by
about one magnitude deeper than the HDF. In this
field, around 10,000 galaxies up to 
$B \sim 30^m$ ($\sim 5 \cdot 10^{-9}$ Jy!) were discovered.

    To improve the impact of the ACS data, the central part
of the HUDF was also observed by the HST with the
NICMOS (Near-Infrared Camera and Multi-Object Spectrometer) with 
filters F110W ($J$) and F160W ($H$). An
unprecedented depth was also attained in these observations: 
the limiting apparent magnitudes in the $J$ and $H$ bands
are $\approx$27.$^m$5 ($\sim 1.5 \cdot 10^{-8}$ Jy) [74]. 
Such high-quality data in
the near-IR spectral range make the HUDF an extremely
valuable field for selecting and studying the most distant
objects. The first analysis of the HUDF allowed discovering
galaxy candidates at $z \sim 7-8$ (see, e.g., [74]).

    In parallel with the main observations, the HUDF field
has been studied with other HST instruments: the STIS
(Space Telescope Imaging Spectrograph) and the WFPC-2
(Wide-Field Planetary Camera 2). These observations and
results are summarized on the STScI site: 
http://www.stsci.edu/hst/udf.

\subsection{Other projects}

Above, we have briefly described typical or unique projects.
This list of notable deep field projects is far from being
complete, but it is impossible to describe all of them. For
completeness, below I briefly enumerate several other interesting 
projects.

\begin{figure}
\centerline{\psfig{file=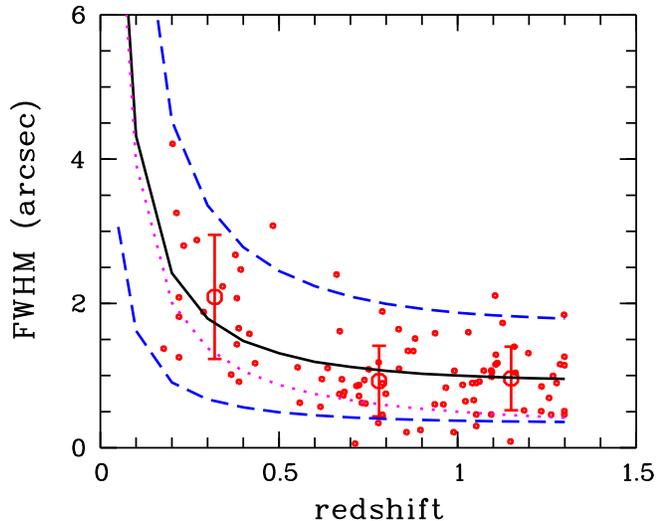,width=8.8cm,angle=-90,clip=}}
\caption{The angular size of bright galaxies as a function of $z$ 
(red points).
The circles with bars show the mean values and corresponding dispersions
for three intervals of $z$. The dashed lines are the lines of constant 
linear sizes (the bottom and upper lines correspond to the respective 
linear sizes 3 kpc and 15 kpc). The solid line shows the dependence 
of the linear size on $z$ for a galaxy with FWHM=8 kpc (FWHM is the full 
width at the half-maximum brightness). The dotted line shows the 
expected change in the angular size of a galaxy according to the 
law  $\propto (1+z)^{-1}$ as predicted by some models.}
\end{figure}

    The LCRS (Las Campanas Redshift Survey) is a spectroscopic 
survey of $\sim$2600 galaxies with the 2.5-m telescope of
the Las Campanas observatory (Chile) [75]. The survey covers
about 700 sq. deg. and consists of six extended patches 
$1.^{\rm o}5\times 80^{\rm o}$ each.

    The NDF (NTT Deep Field) represents a deep ($B \leq 27.^m5$)
photometry of a small ($\sim$5 sq. min.) area with the
3.6-m ESO NTT telescope [76].

    The CNOC2 (Canadian Network for Observational
Cosmology) is a survey covering $\sim$1.5 sq. deg. 
of the sky with the 3.6-m CFHT telescope [77]. The survey is aimed at
determining redshifts for $\sim$6000 galaxies with the apparent
magnitude $R \leq 21.^m5$ and providing multicolor photometry for
$\sim$40,000 galaxies with $R \leq 23^m$.

    The MUNICS (Munich Near-Infrared Cluster Survey)
represents a photometric and spectroscopic study of several
thousand galaxies with $K \leq 19.^m5$ within several areas with
the total coverage $\sim$1 sq. deg. [78] The photometry was obtained
with the 2.2-m and 3.5-m telescopes of the Calar Alto
observatory (Spain). The Hobbey-Eberly (9.2-m, USA) and
ESO-VLT telescopes were also used for spectral observations.

   K20 is a spectroscopic survey of a complete sample of
galaxies with the apparent magnitude $K < 20^m$  (around
550 objects) within two fields covering $\sim$52 sq. min. with 
the ESO VLT telescope [79].

   The DEEP2 (Deep Extragalactic Evolutionary Probe 2) is
a spectroscopic survey covering $\sim$3.5 sq. deg. of the sky with a
multi-object spectrograph on the 10-m Keck-II telescope [80].
Redshifts for $\sim$60,000 remote ($z > 0.7$) galaxies will be
measured.

   GEMS (Galaxy Evolution from Morphologies and
SEDs) represents the largest ($\sim 28' \times 28'$) image obtained
up to the present time with the Hubble Space Telescope [81].
The field was observed with two filters (F606W and F850LP).
The integrated image represents a mosaic from about 60
WFC ACS fields. The GEMS field is centered on the CDF-S
and includes the GOODS field. The size and location of
GEMS are almost identical to the COMBO-17. The GEMS
data allow the study of the structure and morphology of
$\sim$10,000 galaxies.

   The VVDS (VIMOS-VLT Deep Survey) is a photometric
and spectroscopic survey of $\sim$100,000 galaxies within several
deep fields covering $\sim 16$ sq. deg. with the VIsible Multi-Object
Spectrograph (VIMOS) on the ESO VLT telescope [82, 83].

    The OACDF (Capodimonte Deep Field) is a multi-color 
(9 color bands) photometric survey of $\sim$50,000
galaxies within a $\sim$0.5 sq. deg. field with the ESO/MPG 2.2-m
telescope [84].

The NDWFS (NOAO Deep Wide-Field Survey) is a deep optical
and near-infrared imaging survey covering two 9.3 sq. deg.
fields with the KPNO and CTIO telescopes 
(http://www.noao.edu/noao/noaodeep/). 

COSMOS is an HST Treasury project to perform a survey using
the ACS in a single filter (F814W) in a contiguous 2 sq. deg.
equatorial field (http://cosmos.astro.caltech.edu).
The selection of an equatorial field has allowed observatories
in both hemispheres to join efforts for the extensive follow-up
multi-wavelength observations.

    This list is of course incomplete. In particular, there are
many purely spectroscopic galaxy surveys in small fields,
which are referenced in papers cited above.

    The description of the projects shows that virtually all the
largest ground-based telescopes are participating in deep field
studies and surveys. The largest Russian 6-m telescope of
SAO RAN (BTA) is no exception. Shortly after its construction (1976), 
deep galaxy counts near the north galactic pole
were carried out with this instrument [85]. These counts (the
black box in Fig. 2 a,b) were followed up to $B \approx 26^m$, which
was one of the best results at that time.

    The possibilities of the BTA are well illustrated by paper
[86] reporting the results of four-color photometry of a small
($3.'6 \times 3'$ ) field centered at the gamma-ray burst GRB\,000926. 
Around 300 objects with $B \leq 26.^m6$ were found in
this field (the black circle in Figs 2 a,b). Both differential
galaxy counts and their general characteristics turned out to
be in agreement with the results of other deep fields. An
analysis of images of bright ($M(B) < -18^m$, where $M(B)$ is
the absolute magnitude in filter $B$) spiral galaxies led to
the conclusion that there is no strong evolution of their linear
sizes at $z \leq 1$ (Fig. 10, [86]).

\section{Some results}

In this Section, we briefly report selected results of the
described projects. We describe very general {\it observational}
results on the mean characteristics of galaxies.

\subsection{Galaxy counts}

\begin{figure}
\centerline{\psfig{file=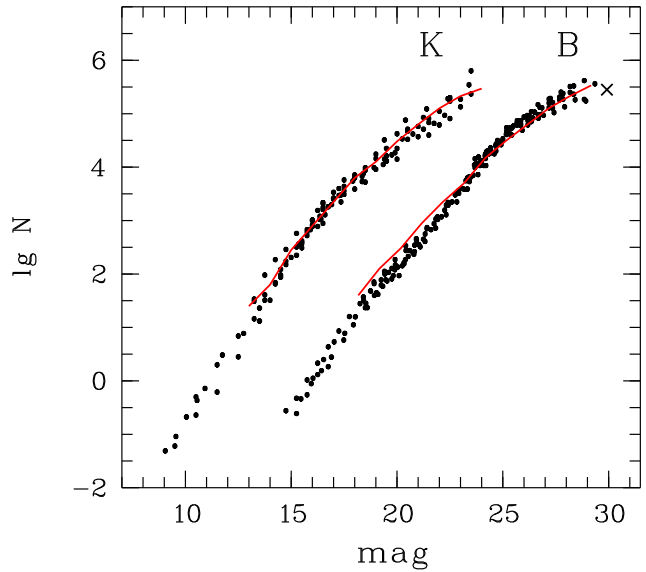,width=8.8cm,angle=-90,clip=}}
\caption{Differential counts of galaxies (the number of galaxies within a
given range of apparent magnitudes normalized to 1 square degree) with
filters $B$ and $K$ (dots). Data in filters $B$ and $K$ are calculated 
using 0.$^m$5 and 1.$^m$0 intervals, respectively. The cross marks 
the HUDF galaxy counts.
The solid lines show predictions of the galaxy formation model in [88].}
\end{figure}

Over many years, galaxy counts, i.e., the plot of the observed
number of galaxies at the limiting magnitude, have
been considered to be an important cosmological test. In
particular, in the 1930s, Hubble tried to apply them to
estimate the curvature of space. It became clear later that
practical application of this test is so difficult (photometric
errors, the account for $k$-correction, the evolution of galaxies
with time) that ``any attempt to do so appears to be a waste of
telescope time'' [87]. Presently, deep counts are regarded not
as a cosmological test but rather as a test of galaxy formation
and evolution.

   Figure 11 summarizes modern results of differential galaxy
counts according to data from the web page 
http://star-www.dur.ac.uk/$\sim$nm/pubhtml/counts/counts.html. 
Only data obtained after 1995 are shown. With each filter, the 
results of about twenty projects (including 2MASS, SDSS, HDF, CDF,
NDF, etc.) are summarized; with filter $B$, counts in the SDF,
VVDS, and HUDF fields are added. The figure shows good
agreement between the results of different works. For
example, for $B\approx25^m$, the count dispersion is only about
10\% (accounting for the photometry, different selection of
galaxies, etc., this dispersion must be even smaller), which
clearly illustrates the homogeneity and isotropy of large-scale
galaxy distribution. For brighter (and, on average, closer)
objects, the count dispersion increases due to large-scale
structure effects. The weakest source counts strongly suffer
from photometric errors and other factors.

    The solid lines in Fig. 11 show predictions of a semianalytic 
model of galaxy formation [88] (the `LC' model in
that paper). Both this and other models (see, e.g., [89] and the
references therein) can satisfactorily fit observations. However, 
the model predictions are not fully definitive due to
many parameters characterizing galactic properties and their
evolution with $z$ (including spatial density evolution). For
further progress in this field, both observational data and
theoretical understanding of galaxy evolution must be
improved.

\subsection{Galaxy distribution}

\begin{figure}
\centering{
\vspace*{-3.0cm}
\hspace*{6cm}
\vbox{
\includegraphics{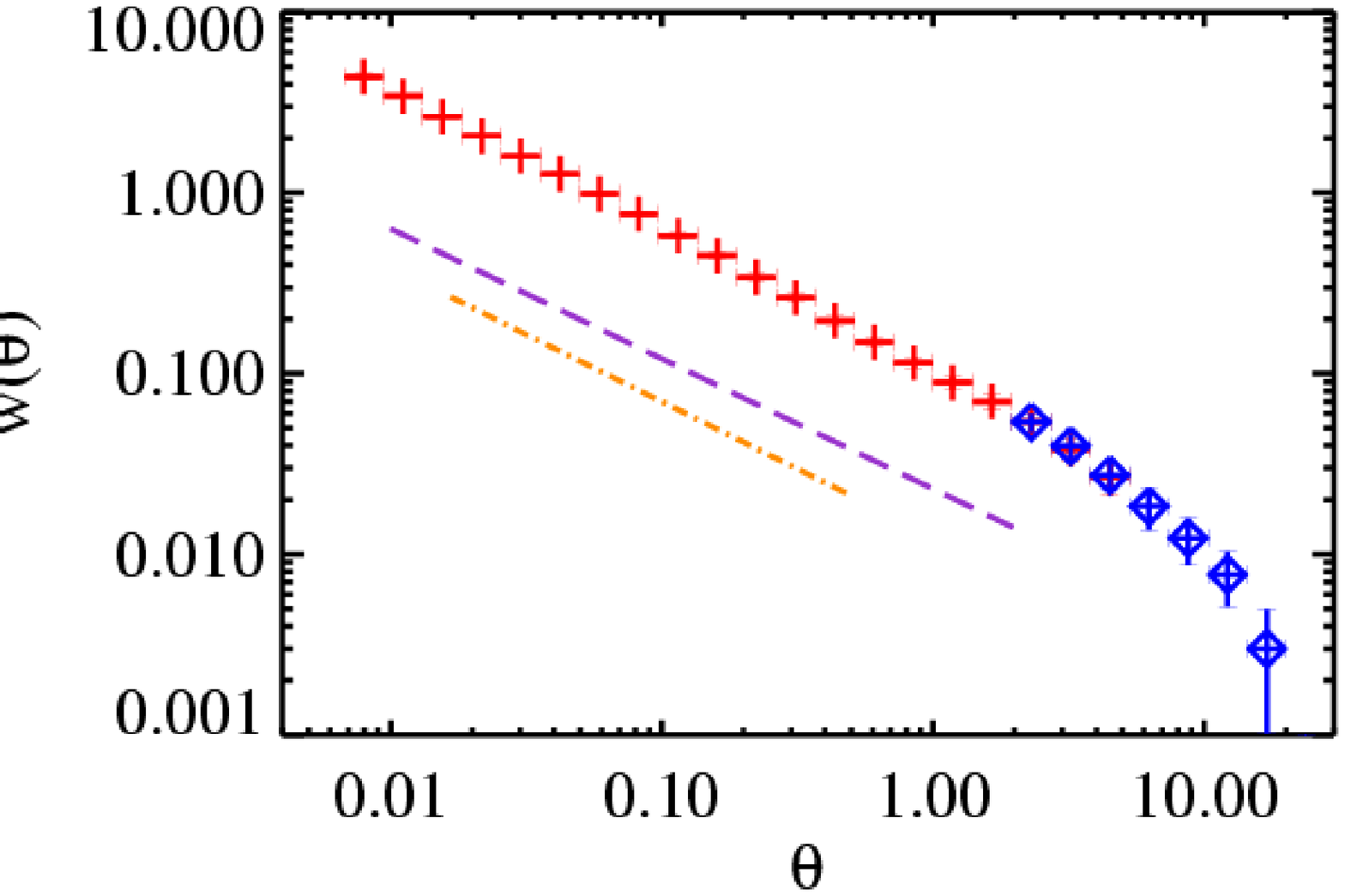}
\includegraphics{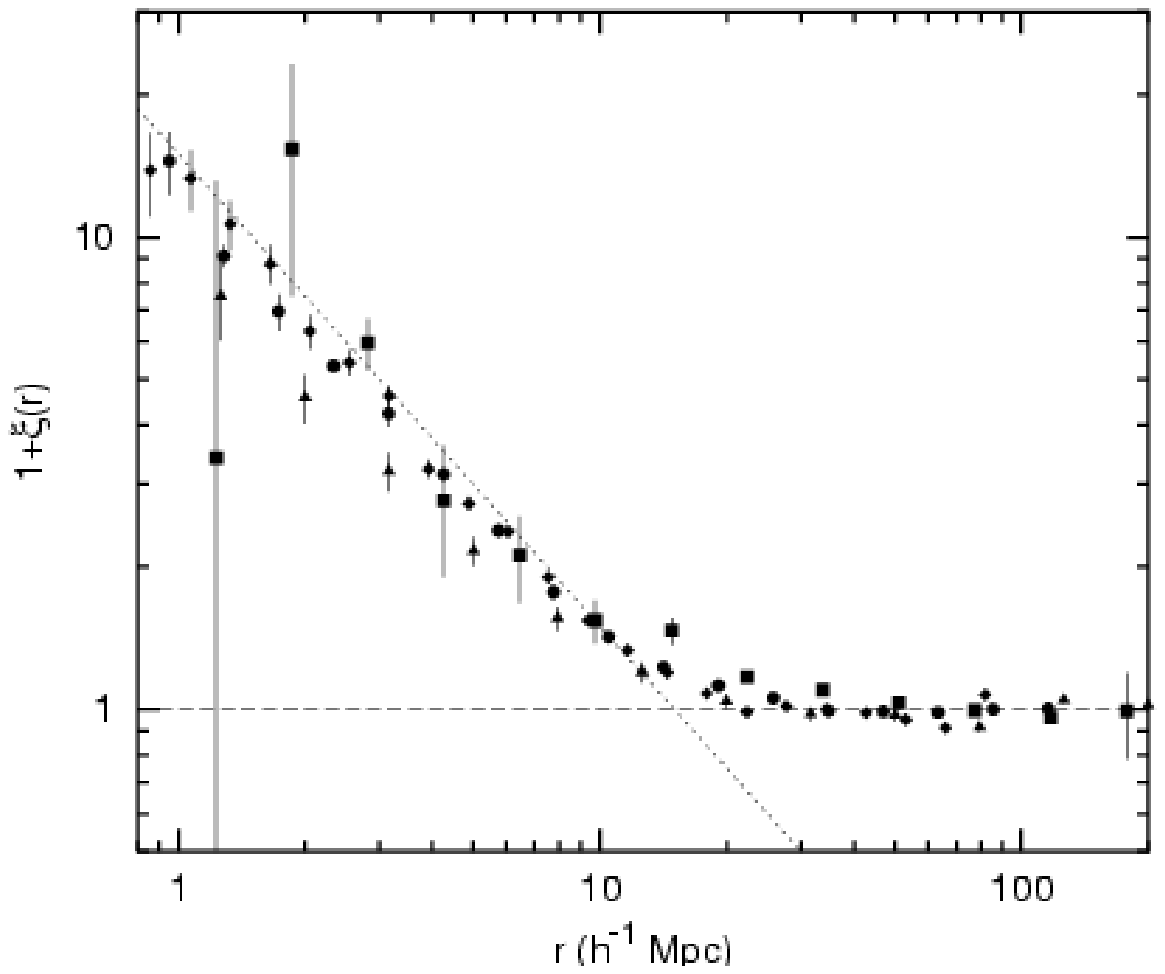}
}\par
\vspace*{16cm}
\caption{Top: the two-point angular correlation function 
$\omega(\theta)$ (the
angle $\theta$ in the plot is measured in degrees) for 2MASS galaxies 
(crosses and diamonds). The dashed and dot-dashed straight lines 
plot linear approximations of the data obtained in the APM and SDSS 
surveys, respectively [91]. Bottom: the two-point correlation function 
$\xi(r)$ according to data from different surveys 
(including APM, LCRS, etc.) [92].}}
\end{figure}

The distribution of galaxies in the nearby volume of the
Universe is highly inhomogeneous (Figs 3, 4). When passing to
the hundred Megaparsec scale, the density fluctuations
smoothen and the distribution becomes more homogeneous
(Fig. 5).

\begin{figure}
\caption{Density fluctuations $\delta\rho/\rho$ as a function of 
scale [94] The solid line marked with CDM shows the prediction of 
a flat CDM model (the model with `cold dark matter'). The `hot dark 
matter' model prediction is marked with HDM.}
\end{figure}

    The clustering of galaxies is usually described in terms of
two-point correlation functions $\xi(r)$ and $\omega(\theta)$. 
The former function describes the joint probability of finding two
galaxies separated by a distance $r$, and the latter characterizes
the joint probability of detecting two objects at the angular
distance $\theta$ [90]. To calculate $\xi(r)$, spatial distances between
galaxies should be known, and in practice it is therefore more
convenient to measure the (angular) two-point correlation
function $\omega(\theta)$. From $\omega(\theta)$, one can then 
estimate $\xi(r)$ because
both functions are related through the Limber integral
equation. If $\xi(r)$ can be represented as a power law
$\xi(r)=(r/r_0)^{-\gamma}$, the angular correlation also takes 
a power-law form $\omega(\theta) \propto \theta^{1-\gamma}$ [90].

    Figure 12 plots the angular correlation function for
$\sim$0.5 million galaxies from the 2MASS survey [91]. At
angular scales $1' < \theta < 2.^{\rm o}5$, this function is well 
fit by a power law with $1-\gamma$=--0.79$\pm$0.02. The amplitude 
of $\omega(\theta)$ depends on the sample depth -- for brighter 
and closer objects, the clustering amplitude increases (this, 
in particular, explains the systematic shift between different 
survey data in Fig. 12).

    In the bottom part of Fig. 12, we show the correlation
function $\xi(r)$ calculated in different papers (h in the figure
means the Hubble constant value expressed in units of
100 km/s/Mpc). In the range 0.1 Mpc $\leq  r \leq 20$ Mpc,
this function follows a power law with the exponent
$\gamma \approx 1.7-1.8$, and then tends to zero. The characteristic
clustering scale (correlation length) $r_0$ for nearby galaxies is
$\approx$7 Mpc. The correlation length depends on the properties of
galaxies, such as their luminosity and morphological type
(see, e.g., [93]), but is independent of the sample depth (see the
discussion in [92]).

    Modern survey data allow determining the density
fluctuations in the Universe as a function of the scale of
averaging (Fig. 13). (The power spectrum of the SDSS
galaxies, which has been used to plot this figure, is based on
data on 2$\cdot$10$^5$ galaxies [95].) Figure 13 shows that 
different kinds of data, from galaxy density fluctuations to cosmic
microwave background anisotropy, form a unique smooth
dependence described by the CDM model.

     Recently completed surveys have allowed features of the
nearby galaxy distribution to be studied in a detail unavailable 
earlier. In particular, the 2MASS survey enabled
examining the galaxy distribution in the `zone of avoidance'
Ð the strip near the Milky Way plane ($|b|<10^{\rm o}$) where the
interstellar absorption screens extragalactic objects [96]. The
galaxy distribution as derived from the 2MASS, 2dF, and
APM surveys led to the conclusion that there is a 30\% deficit
of bright galaxies in the southern galactic hemisphere in
comparison with the northern one [97]. The authors believe
that the observed deficit is a consequence of a huge `hole' with
a linear size possibly exceeding 200 Mpc in the local galaxy
distribution. Such a large local nonhomogeneity, as well as
the possible presence of a well-defined large-scale structure at
$z\sim6$ ([66], see Section 4.5; [98]), can pose certain problems
for the standard CDM model. We note that nonhomogeneities 
of a comparable scale ($\geq$200 Mpc) have been found in
the quasar distribution derived from the 2QZ project (see
Section 3.5) [99].

\subsection{Evolution of the luminosity function}

The luminosity function (LF) is the dependence of the
number of galaxies within a unit volume on their luminosity.
It is one of the most important integral characteristics of
galaxies. The LF allows estimating the mean luminosity
density in the Universe. The LF form is one of the main tests
of galaxy formation models. The standard form of the LF is
the so-called Schechter function [100] 

$\phi(L)dL\,=\,\phi_*\,(\frac{L}{L_*})^{\alpha}\,\exp(-\frac{L}{L_*})\,d(\frac{L}{L_*})$,\\
where $\phi(L)dL$ is the number of galaxies with the luminosity
from $L$ to $L+dL$ per unit volume, and $\phi_*$, $L_*$ and 
$\alpha$ are parameters. The parameter $\phi_*$ yields the 
normalization of the LF, $L_*$ is the characteristic luminosity, 
and $\alpha$ determines
the slope of the weak wing ($L < L_*$) of the LF: the weak wing
of the LF is flat for $\alpha=-1$, the LF increases with decreasing
$L$ for $\alpha < -1$, and decreases at $\alpha > -1$. The Schechter
function fits well the real LF of field galaxies and clusters
and has convenient analytical properties.

   At present, the local LF of galaxies is relatively
well studied. According to many papers (including the 2dF
and SDSS surveys), within the absolute magnitude range
$-15^m \geq M(B) \geq -22^m$ (filter $B$), the LF can be described
with the following values of parameters: \\
$\alpha\approx-(1.1-1.2)$, \\
$M_*(B)\approx-20.^m2$ ($L_*(B)=1.9 \cdot 10^{10}\,L_{\odot,B}$), and \\
$\phi_*\approx(0.5-0.7)\cdot10^{-2}$\,Mpc$^{-3}$
(see. e.g., [101]).\\ Therefore, the luminosity density of galaxies 
at $z=0$ is \\
$\rho_L(B) = \phi_* L_* \Gamma(\alpha+2) \approx  1.3 \cdot 10^8$
$L_{\odot,B}$/Mpc$^3$ \\and the galaxy density is\\
$\rho = \rho_L/L_* = \phi_* \Gamma(\alpha+2) \sim 10^{-2}$ Mpc$^{-3}$.\\ 
The LF of local galaxies depends on their morphological type
and environment [102].

   Numerous deep field studies performed over the last ten
years have enabled the evolution of the LF with $z$ to be
determined. In solving this problem, the so-called `photometric 
redshifts' inferred from multicolor photometry are
used instead of spectroscopic ones for the most distant
objects. Such a photometry allows a kind of a low-resolution
spectrum and hence $z$ of an object to be obtained. Photometric 
estimations of $z$ are being made with $\approx$10\%--20\%
accuracy, which is quite sufficient to derive the LF for large
samples of galaxies.

\begin{figure}
\centering{
\vspace*{-3cm}
\hspace*{6cm}
\vbox{
\includegraphics{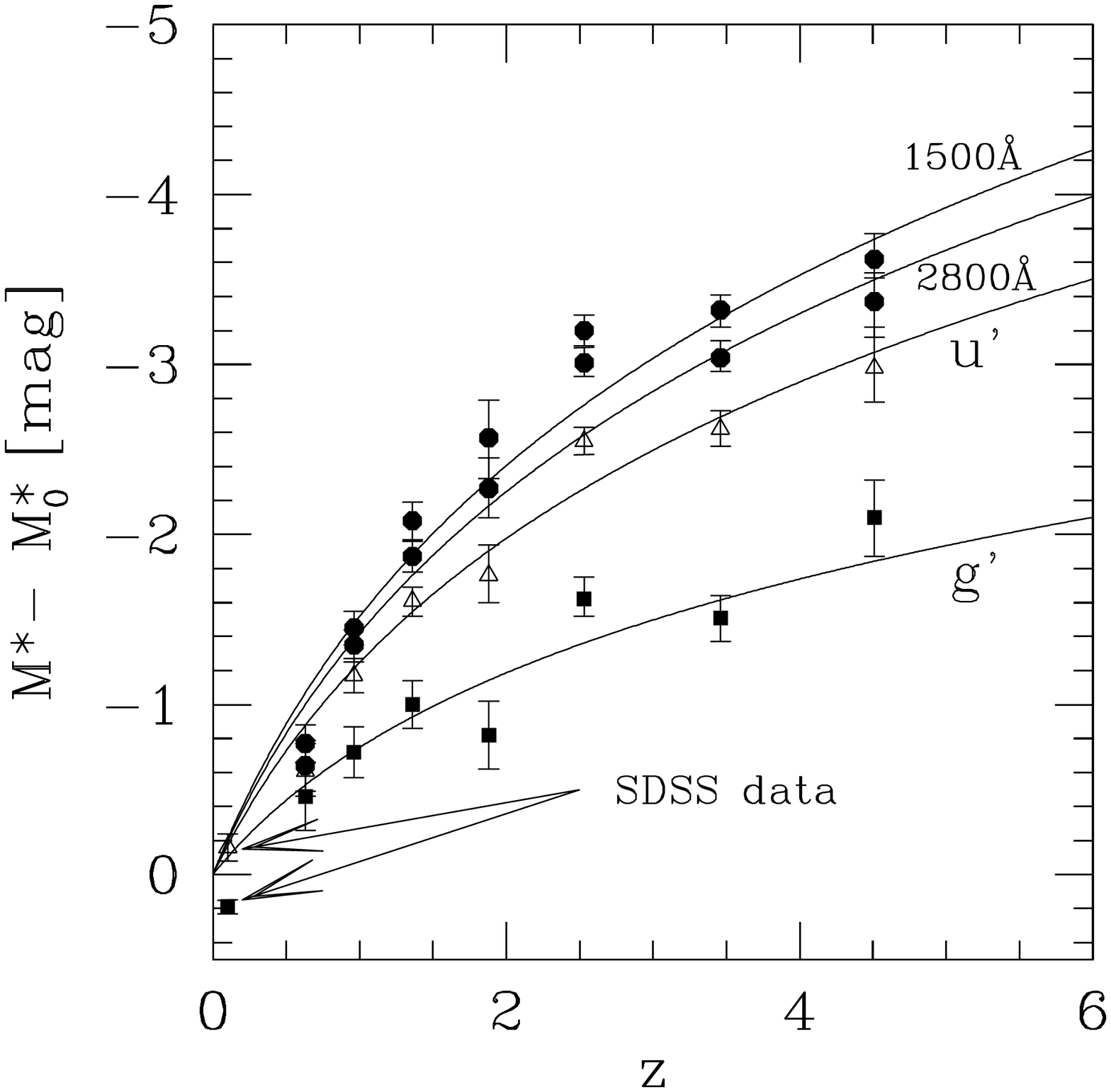}
\includegraphics{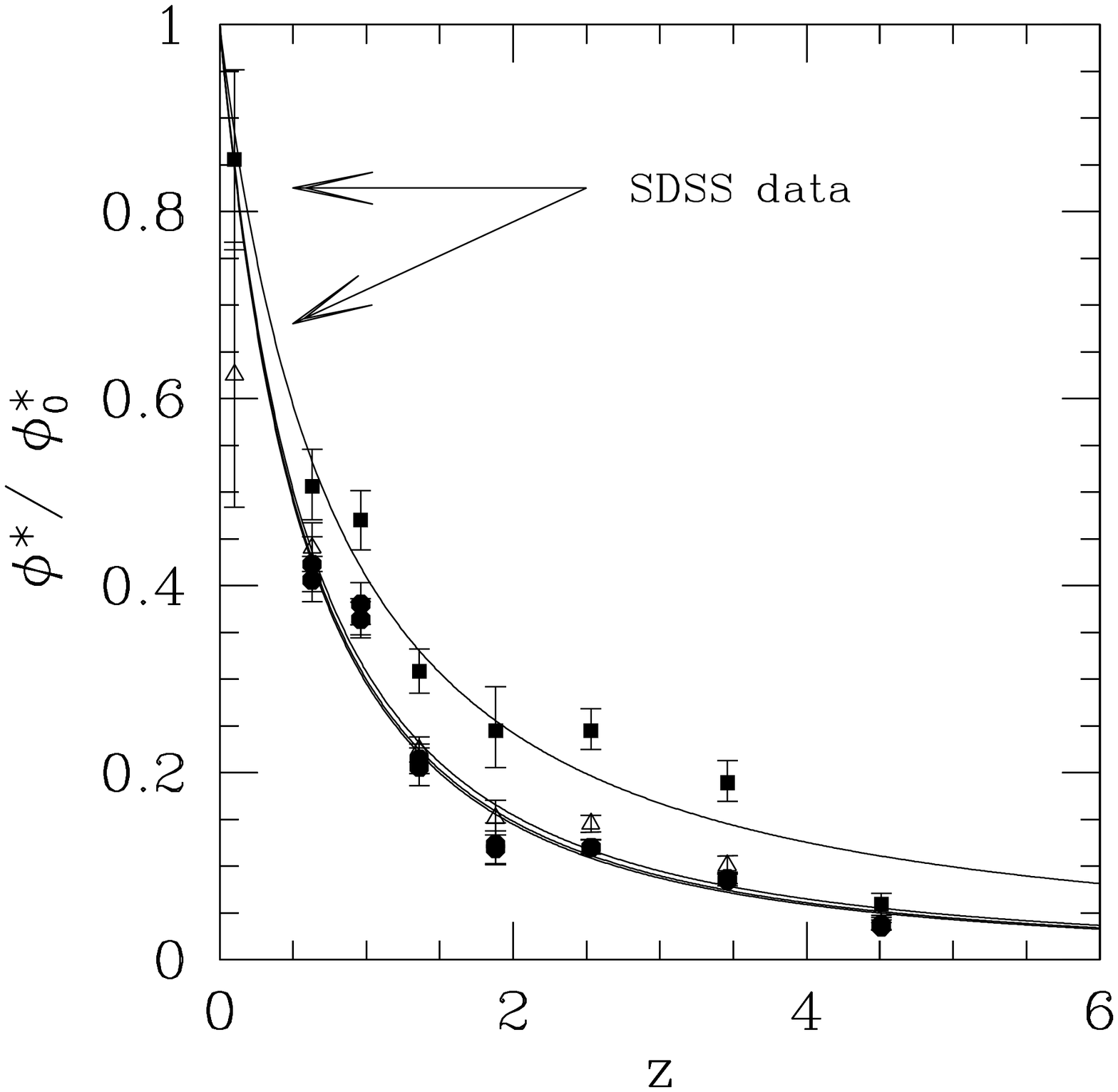}
}\par
\vspace*{19cm}
\caption{The redshift dependence of the parameters of the luminosity
function of galaxies $M_*$ (top) and $\phi_*$ (bottom) [103]. Different 
color bands are marked with different signs. The solid lines show 
analytic approximations of the observational data. The LF parameters 
as derived from the SDSS survey are shown by arrows.}
}
\end{figure}

    Observations suggest a differential (depending on the
galaxy type and the color band) evolution of the LF.
Different papers give somewhat different results, but the
qualitative picture emerging is as follows: the value of $M_*$
increases with $z$, while $\phi_*$ decreases (Fig. 14). According to
[103], towards $z\sim5$, the value of $M_*$ in filter $B$ increases by
1$^m$--2$^m$, and $\phi_*$ decreases by 5--10 times. The evolution of the
LF slope is much less definitive, although some authors note a
decrease in $\alpha$ with $z$. By considering different types of objects
separately, the space density of elliptical and early spiral
galaxies almost stays constant or slightly decreases toward
$z\sim1$, while their LF evolution can be described as a change in
the luminosity of galaxies (they become brighter). In contrast,
the space density of late spiral galaxies with active star
formations notably increases toward $z\sim1$ [104, 105]. The
change in the LF of galaxies alters the luminosity density they
produce: from $z=0$ to $z\sim3$, the value of $\rho_L$ increases, with
the strongest growth being in the UV region (by about 5 times
[106]).

\subsection{Evolution of the galaxy structure}

One of the main goals of the deep field galaxy studies is the
origin and evolution of the Hubble sequence. In the local
Universe, the optical morphology of the vast majority of
bright galaxies can be described in terms of a simple
classification scheme suggested by Hubble [3]. Only about
5\% of nearby objects do not fit this scheme and are related to
irregular or interacting galaxies [107, 108].

    The deep HST fields for the first time allowed us to see the
structure of distant galaxies. The very first studies revealed
that the fraction of galaxies that do not fit the Hubble scheme
increases for fainter objects [109]. At $z\sim1$ (where the age of
the Universe is about half the Hubble time), the fraction of
such galaxies reaches 30\%--40\% (see examples in Figs 6 and 9).
The lack of a convenient classification for distant galaxies
stimulated the development of new methods for analyzing
their images and the construction of objective classification
schemes invoking the characteristics such as asymmetry and
concentration indices (see, e.g., [109, 110]).

    The statistics of objects in some deep fields also suggests
that the fraction of interacting galaxies and merging galaxies
increases with $z$. With the $(1+z)^m$ growth assumed, observational 
data suggest $m\approx2-4$ for $z\sim1$ [111, 112]. The
evolution of the merging rate is likely to depend on the
mass of galaxies -- it is most pronounced for massive
objects [112].

    It is much more difficult to draw definitive conclusions on
the structure evolution for objects at $z \geq 1$ due to the
increasing effects of the $k$-correction, the cosmological
diming of surface brightness and degradation of the 
resolution [113]. The history with bar studies for distant galaxies
may serve as an instructive illustration. The first morphological 
studies of galaxies in the HDF implied a drastic decrease
in the fraction of barred spirals at $z > 0.5$, but the subsequent
analysis indicated that this fraction remains almost constant
($\sim$40\%), at least up to $z=1.1$ [114, 115].

     A significant amount of observational data has been
obtained about changes in the characteristics of large-scale
subsystems of galaxies. For example, it has been established
that by $z\sim1$, the disk surface brightness of spiral galaxies
increases by $\sim1^m$, while the color indices decrease (i.e.,
become `bluer') [116, 117]. For $z\leq1$, a change in the slope of
the Tully--Fisher relation is found (the Tully--Fisher relation
is the dependence between the maximum rotation velocity and
the luminosity of spiral galaxies) [118]. This slope change is
believed to be a consequence of the differential evolution of
spiral galaxies: at $z\sim1$, low-mass spirals become brighter by
1$^m$--2$^m$, while massive ones stay virtually as bright. Disks of
`edge-on' spirals at $z\sim1$ show a larger relative thickness (the
ratio of the vertical and radial scales in the brightness
distribution) and demonstrate vertical deformations of the
plane (warps) more frequently than nearby objects [119, 120].
Spectral studies of spiral galaxies suggest their chemical
composition evolution: from $z=0$ to $z=1$, the metallicity
of gas subsystems of galaxies decreases [121].

    Some papers have investigated photometric and kinematical 
characteristics of elliptical galaxies in the field and in 
clusters up to $z\sim1$ (see, e.g., [122, 123]). A deviation of 
distant early-type galaxies from the Fundamental Plane determined 
by the nearby objects has been discovered. This deviation is
explained by the so-called `passive' evolution of their
luminosities and, correspondingly, the mass--luminosity
ratio (see [40, 124] for more details).

\subsection{The most distant galaxies}

\begin{figure}
\centerline{\psfig{file=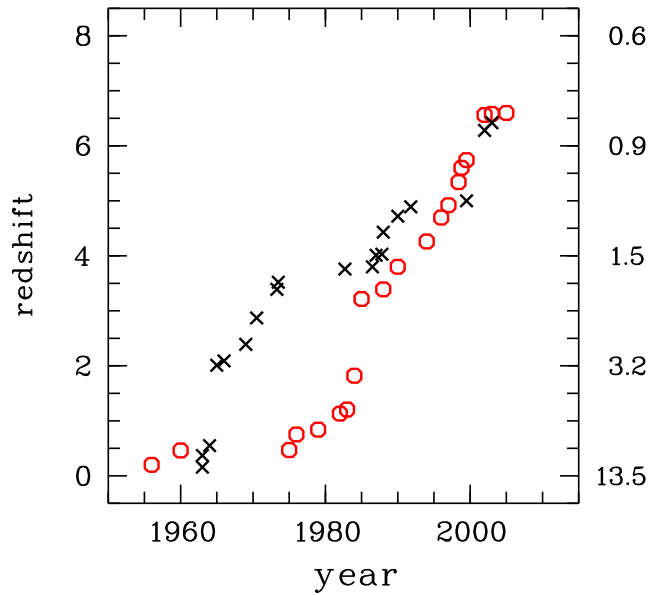,width=8.8cm,angle=-90,clip=}}
\caption{The history of detection of the most distant objects in the
Universe (the circles are for galaxies, the crosses are for quasars). 
Years when the objects were discovered are plotted along the horizontal 
axis. Along the vertical axis to the left are plotted redshifts, and to 
the right -- time in billions of years since the beginning of the 
cosmological expansion. The plot relies on the data in review [125] 
added with data obtained in recent years.}
\end{figure}

Searches for and studies of the most distant galaxies in the
Universe are some of the most interesting and important fields
of extragalactic astronomy. The most distant and hence the
youngest galaxies provide invaluable tests of galaxy formation 
models and allow processes in the relatively early
Universe to be studied.

     The history of discoveries of the most distant galaxies is
shown in Fig. 15. It is seen that quasars had been the most
distant objects over almost three decades. (The term `quasar'
itself had often served as a synonym for the most distant
objects.) The explanation is simple, because quasars are
associated, as a rule, with very bright galaxies whose spectra
show powerful and wide emission lines. The brightness of
quasars and lines in their spectra make them much easier to
observe from cosmological distances than ordinary galaxies
(for example, see the spectrum of a distant quasar in Fig. 16).
At the beginning of the new millennium, new methods
appeared that allowed a very effective selection of ordinary
galaxies at high $z$; since then, these galaxies and not quasars
have been the most distant known objects in the Universe
(Fig. 15).

  There are several methods of selecting galaxies at high $z$.
Analysis of broad-band color indices to find galaxies with
peculiar colors (see Section 2) is one of the most effective
means of selecting very distant galaxies. This method is
primarily aimed at searching for galaxies with an energy
distribution break near the Lyman continuum (912 \AA),
which is expected in star-forming galaxies [9]. Due to the
absorption in L$\alpha$ clouds along the line of sight, emissions in
the spectra of distant galaxies between 912 \AA\, and the 
L$\alpha$ line
are absorbed. This creates an additional spectral feature that
allows distant galaxies to be selected by their broad-band
color indices. More than a thousand objects with $z > 2.5$ have
been selected using this method; they are commonly referred
to as Lyman-break galaxies or simply LBGs (see [9] for a
detailed review).

\begin{figure}
\centering{
\vspace*{-2.5cm}
\hspace*{7cm}
\vbox{
\includegraphics{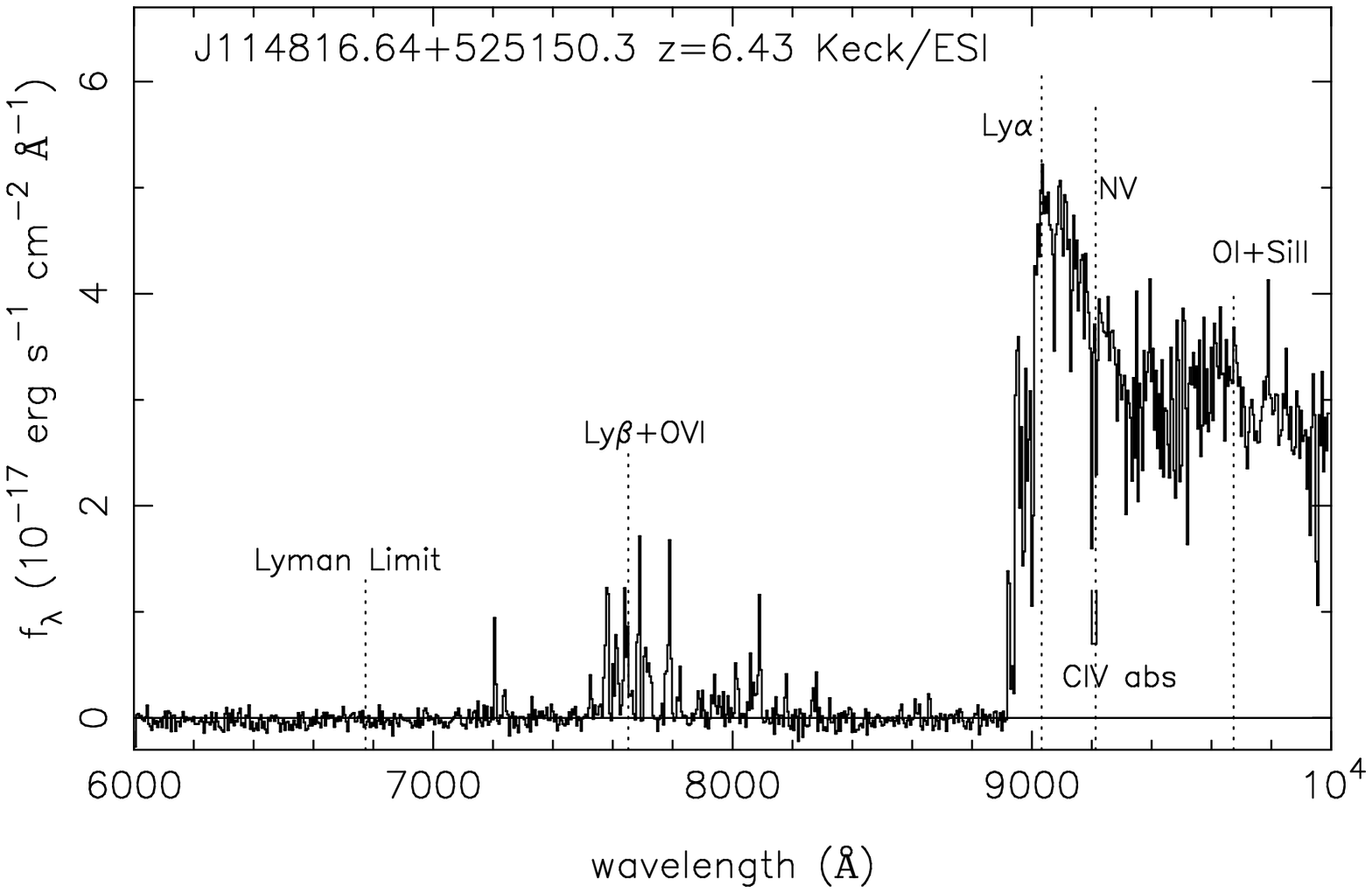}
\includegraphics{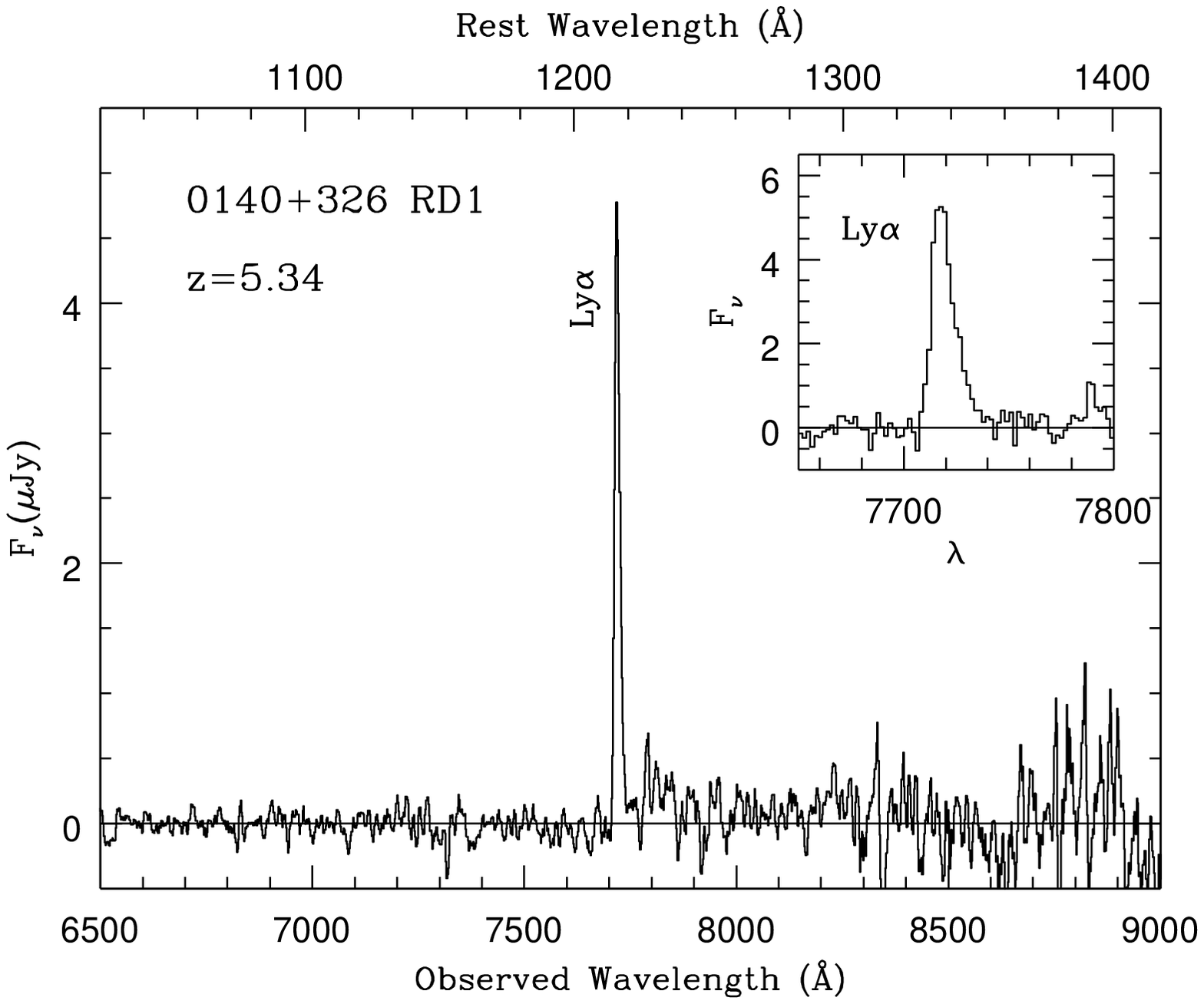}
}\par
\vspace*{16.6cm}
\caption{The spectrum of a quasar with $z=6.4$ (top) [126] and the
spectrum of a galaxy with $z=5.34$ (bottom) [127].}
}
\end{figure}

    The second method frequently used is the search for
galaxies showing a strong emission L$\alpha$ line using a deep
narrow-band photometry of individual areas of the sky
followed by a spectroscopic study of the detected objects
(see Section 2 for more details). Objects found this way are
often called `L$\alpha$ emitters,' or LAEs. It is by using this method
that the most distant object known so far with $z=6.60$ was
discovered and signs of the presence of a large-scale structure
of galaxies at $z=5.7$ were found (Section 4.5).

    At present, more than thirty galaxies with spectroscopic
$z > 5$ have been reliably detected [64, 128] (see an example of
the spectrum in Fig. 16). The age of such objects does not
exceed $\sim$10\% of the age of the Universe (Fig. 15). Distant
galaxies have been discovered both in deep fields and near
galaxy clusters enhancing fluxes from remote background
galaxies due to gravitational lensing.

    The main observational features of galaxies with $z > 5$
(see, e.g., [64, 129]) are as follows:\\
    -- as a rule, morphologically peculiar, asymmetric,
compact shapes (the characteristic linear size is 1--5 kpc);\\
    -- a very high surface brightness and luminosity
(corrected for the cosmological brightness decrease and
$k$-correction effects); \\
    -- equivalent widths of the L$\alpha$ line in the comoving frame
are W(L$\alpha$)$\sim$20--100\,\AA;\\
    -- the star formation rate inferred from the L$\alpha$ line
luminosity is $\sim$5--10\,M$_{\odot}$/yr (this rate estimated from
the UV continuum luminosity is several times larger).

    These characteristics are very strongly biased by the
selection procedure itself, and it is therefore unclear to what
extent they reflect actual properties of all objects located at
$z > 5$. The observed objects can be `building blocks' that later
merge and accrete the surrounding matter to form the
galaxies we now know in our vicinity. On the other hand,
some of these objects can represent bulges of massive spirals
under formation or elliptical galaxies.

    The clustering of LBGs and LAEs has been studied in
some papers. For bright galaxies ($L \geq L_*$), the clustering scale
$r_0$ does not show significant evolution from $z=0$ to $z=5$
[130, 131]. In contrast, the `bias parameter' $b$ characterizing
the difference in space distribution of galaxies and dark halos
increases several times toward $z=5$ [131]. The GOODS and
HUDF results may indicate an evolution in galaxy sizes: from
$z\sim2$ to $z\sim6$, the mean linear sizes decrease by about two
times [132, 133]. Both the galaxy clustering evolution and
change in the galaxy sizes can be explained by the CDM
model of galaxy formation.

    The spectra of the most distant galaxies and quasars
provide the possibility of studying an early evolution of the
intergalactic medium. In particular, the so-called Gunn--Peterson 
effect [134] (a trough in the spectra of distant
objects shorter than L$\alpha$ due to absorption by neutral
hydrogen clouds along the line of sight) allows estimating the
redshift at which the secondary ionization epoch (re-ionization) 
of the Universe has been completed [135]. The discovery
of this effect in spectra of quasars with $z > 6$ (Fig. 16) and its
absence for objects with $z \leq 6$ suggest the re-ionization epoch
(i.e., ionization of the intergalactic medium by radiation from
star formation regions and active galactic nuclei) to have been
completed by $z \sim 6$ [136, 126]. On the other hand, the cosmic
microwave background anisotropy measurements may evidence the 
beginning of secondary ionization at $z \sim 20$ (see,
e.g., review [137]). The combination of these data has led to
the conclusion of a complicated, possibly two-stage, history
of the secondary ionization of the intergalactic medium [138].

\subsection{History of star formation in the Universe}

Reconstruction of the global history of star formation in the
Universe from $z \sim 6$ until now appears to be one of the most
intriguing results derived in recent years from sky surveys and
deep fields. Quantitative results by different authors are
somewhat different, but the general trend of star formation
in the unitary comoving volume as a function of redshift,
which is often referred to as the `Madau diagram/plot' [139],
is likely to be firmly established (see, e.g., Fig. 17).

\begin{figure}
\centerline{\psfig{file=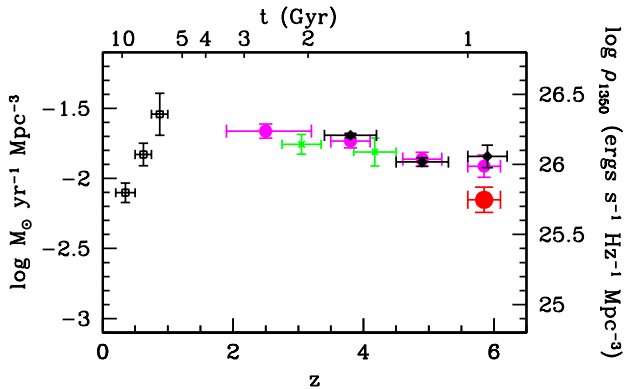,width=8.8cm,clip=}}
\caption{The history of star formation in the Universe [140]. The specific
star formation rate in units M$_{\odot}$/yr/Mpc$^3$ is plotted to the 
left, the luminosity density at $\lambda=1350$\,\AA\,\, is plotted to the 
right. The upper horizontal axis plots the time since the beginning 
of the cosmological expansion.}
\end{figure}

    There are two approaches to constructing this plot. The
first relates to a detailed modeling of the star formation
history in nearby galaxies using their spectra. The second is
more direct and assumes studies of complete samples of
galaxies observed at different $z$. The main problems of this
method are relatively small samples of distant galaxies (which
is related to the small sizes of deep fields) and poorly known
correction for the intrinsic absorption, which can notably
reduce the observed luminosity of distant objects. Nevertheless, 
both approaches yield generally consistent results
(see, e.g., [141]).

    As seen in Fig. 17, the specific star formation rate rapidly
grows from $z=0$ to $z\sim1$, has a global maximum at $z\sim1-2$,
and then starts decreasing, remaining, however, significant up
to the limiting redshift $z$ ($\sim6$) at which modern estimates are
possible. Analysis of the history of star formation in the
Universe leads to the conclusion that 50\% of all stars existing
at $z=0$ were born at $z \geq 1$, $\sim$25\% appeared at $z \geq 2$, 
$\sim$15\% appeared at $z \geq 3$, and $\sim$5\% existed already 
at $z=5$ [142].
One more important observational result is that the number
of massive star systems (with masses exceeding 10$^{11}$\,M$_{\odot}$)
decreases with $z$, although a small number of such galaxies
are also present at $z > 4$ [142].

\section{Conclusion}

Hundreds of years ago, we inhabited a dwarf Universe
consisting of planets and limited by a sphere of fixed stars
(Fig. 1). One hundred years ago, the concept of the Universe
consisting entirely of stars dominated. That Universe
appeared as a huge oblate star cluster, the Milky Way,
with the Sun located close to its center. At the beginning of
the third millennium, the volume of the Universe available
to astronomical observations includes hundreds of billions
of `recessing' galaxies (Fig. 18\footnote{It is instructive to 
note a certain visual similarity between Fig. 18 and
Fig. 1.}), surrounded by dark matter halos and embedded in an 
anti-gravitating cosmic vacuum [144]. Our Universe has already 
passed its peak activity and we are living in its history of 
`decline' when the epoch of active star formation has already 
ended and galaxies have recessed from each other and only 
relatively rarely interact with each other.

\begin{figure*}
\caption{The map of the Universe [143]. An equatorial slice 
($-2^{\rm o}<\delta<2^{\rm o}$) of the galaxy distribution in the 
sky is presented. The right ascension of objects is plotted along 
the horizontal axis. The left and right vertical axes show the 
distance measured in Mpc and in Earth radii, respectively.}
\end{figure*}

     The sky surveys and selected deep field studies, some of
which have been discussed in this review, play a very
important role in creating a large-scale picture of the world.
Modern sky surveys provide information on the characteristics and 
spatial distribution of millions of galaxies. Deep
fields allow galaxies under formation and their evolution over
billions of years to be studied. According to some researchers,
should the current progress in observations be continued, the
question of the formation and evolution of galaxies will be
answered in one or two decades. At that time, we might ``look
back to this one and recall with nostalgia how exciting it was
to help write that story'' [145].

     The wealth of observational data obtained by the projects
described in this review has already been counted in dozens of
terabytes. These data, as a rule, are freely accessible and can be
used through the world-wide web by both professional
researchers and amateurs. All major ground-based telescopes 
and space observatories also have data archives
where results of observations are stored (the BTA observations 
archive can be found at http://www.sao.ru).

     The huge available amount of information on the Universe
has altered the face of modern astronomy. The current rate of
observational data storage on astronomy has reached
$\sim$1 TB/day [146]. Many important tasks, from statistical
studies of the Milky Way and studies of the large-scale
distribution of galaxies to the discovery of new types of
objects, can now be solved without additional observations
by telescopes. These new possibilities, which often escape the
attention of even professional astronomers, pose `technological' 
problems, such as rapid and convenient access to world-wide 
dispersed and often inhomogeneous data, their visualization and 
analysis, etc. This has stimulated the appearance
of the Virtual Observatory concept (http://www.ivoa.net)
primarily aimed at increasing the efficiency of astronomical
studies under conditions of a colossal amount of information [146]. 
Modern and future multi-wavelength sky surveys
create a kind of `virtual Universe' in our computers, and the
Virtual Observatory may turn out to be the principal tool for
its investigation.

     This work is supported by the RFBR grant 03-02-17152.

{\small

{}
}

\small Translated by K.A.\,Postnov; edited by A.M.\,Semikhatov


\begin{thebibliography}{}
\baselineskip=12pt

\bibitem{} 1. Heninger S.K. {\it The Cosmographical Glass. Renaissance
Diagrams of the Universe} (San Marino, California: The Huntington Library,
1977)
\bibitem{} 2. Eremeeva A.I. {\it Vselennaya Gershelya} (Herschel's Universe) 
(Moscow: Nauka, 1966) 
\bibitem{} 3. Hubble E. {\it Astrophys. J.} {\bf 64} 321 (1926)
\bibitem{} 4. Hubble E. {\it Astrophys. J.} {\bf 79} 8 (1934)
\bibitem{} 5. Sandage A., in {\it The Hubble Deep Field} (Eds M Livio,
S M Fall, P Madau) (New York: Cambridge University Press, 1998) p.1 
\bibitem{} 6. Ellis R., in {\it Galaxies at high redshift. XI Canary Islands            
Winter School of Astrophysics} (Eds I Pirez-Fournon, M Balcells, 
F Moreno-Insertis, F Sanchez ) (Cambridge: Cambridge Univ. Press, 2003) 
p.1
\bibitem{} 7. Hubble E. {\it Astrophys. J.} {\bf 84} 517 (1936)
\bibitem{} 8. Rosati P., in {\it Modern Cosmology} (Eds S Bonometto,
V Gorini, U Moschella) {\it Bristol: Institute of Physics Publishing, 
2002}, p.312
\bibitem{} 9. Giavalisco M. {\it Ann. Rev. Astron. Astrophys.} {\bf 40} 579
(2002)
\bibitem{} 10. Markaryan B.E., Lipovetskii V.A., Stepanyan J.A. 
{\it Astrofizika} {\bf 15} 549 (1979)
\bibitem{} 11. Morgan D.H., in {\it The Future Utilisation of Schmidt
Telescopes}, {\it ASP Conference Series} {\bf 84} 137 (1995) 
\bibitem{} 12. Zwicky F., Herzog E., Wild P., Karpowicz M., Kowal C.
{\it Catalogue of Galaxies and of Clusters of Galaxies: I--VI}
(Pasadena: California Institute of Technology, 1961-1968)
\bibitem{} 13. Vorontsov-Velyaminov B.A., Krasnogorskaya A.A.,
Arhipova V.P. {\it Morfologicheskii Katalog Galaktik: I--IV} 
(Morphological Catalog of Galaxies) (Moscow: Izd. MGU, 1962-1968)
\bibitem{} 14. Vorontsov-Velyaminov B.A. 
{\it Atlas i Katalog Vzaimodeistvuyushchih Galaktik} (Atlas and Catalog
of Interacting Galaxies) (Moscow: Izd. MGU, 1959)
\bibitem{} 15. Karachentsev I.D. {\it Dvoynye Galaktiki} (Binary Galaxies)
(Moscow: Nauka, 1987)
\bibitem{} 16. Abell G. {\it Astrophys. J. Suppl.} {\bf 3} 211 (1958)
\bibitem{} 17. Reid I.N., Brewer C., Brucato R.J. et al., {\it Publ. Astron.
Soc. Pacif.} {\bf 103} 661 (1991) 
\bibitem{} 18. Maddox S.J., Sutherland W.J., Efstathiou G., Loveday J.,
{\it Mon. Not. R. Astron. Soc.} {\bf 243} 692 (1990)
\bibitem{} 19. Maddox S.J., Efstathiou G., Sutherland W.J., Loveday J.,
{\it Mon. Not. R. Astron. Soc.} {\bf 242} 43P (1990)
\bibitem{} 20. Dalton G.B., Croft R.A.C., Efstathiou G. et al.,
{\it Mon. Not. R. Astron. Soc.} {\bf 271} L47 (1994)
\bibitem{} 21. Pennington R.L., Humphreys R.M., Odewahn S.C. et al.,
{\it Publ. Astron. Soc. Pacif.} {\bf 105} 521 (1993)
\bibitem{} 22. Yentis D.J., Cruddace R.G., Gursky H. et al.,
{\it Astron. Astrophys. Sci. Lib.} {\bf 174} 67 (1992)
\bibitem{} 23. Hambly N.C., MacGillivray H.T., Read M.A. et al.,
{\it Mon. Not. R. Astron. Soc.} {\bf 326} 1279 (2001)
\bibitem{} 24. Lasker B.M., McLean B.J., Shara M.M. et al., 
{\it Bull. Inform. CDS} {\bf 37} 15 (1989)
\bibitem{} 25. Postman M., {\it Publ. Astron. Soc. Pacif.} {\bf 106} 108 (1994)
\bibitem{} 26. Lasker B.M., Doggett J., McLean B. et al., 
{\it ASP Conference Series} {\bf 101} 88 (1996)
\bibitem{} 27. Djorgovski S.G., Gal R.R., Odewahn S.C. et al., in
{\it Wide Field Surveys in Cosmology} (Eds S Colombi, Y Mellier)
{\it Gif sur Yvette: Editions Frontiers, 1999}, p.89
\bibitem{} 28. Skrutskie M.F., Schneider S.E., Stiening R. et al.,
in {\it The Impact of Large Scale Near-IR Sky Surveys} (Eds F Garzon
et al.) {\it Dordrecht: Kluwer, 1997}, p.25
\bibitem{} 29. Jarrett T.H., {\it Publ. Astron. Soc. Pacif.} {\bf 112}
1008 (2000)
\bibitem{} 30. Jarrett T.H., Chester T., Cutri R. et al., {\it Astron. J.} 
{\bf 119} 2498 (2000)
\bibitem{} 31. Jarrett T., {\it Publ. Astron. Soc. Austr.} {\bf 21} 396
(2004)
\bibitem{} 32. Eichhorn G., {\it Astron. Geophys.} {\bf 45} 3.7 (2004)
\bibitem{} 33. Colless M., Dalton G., Maddox S. et al., 
{\it Mon. Not. R. Astron. Soc.} {\bf 328} 1039 (2001)
\bibitem{} 34. Colless M., Peterson B.A., Jackson C. et al., The 2dF
Galaxy Redshift Survey: Final Data Release, 2003; astro-ph/0306581
\bibitem{} 35. Colless M., in {\it Measuring and Modeling the Universe}
(Ed W L Freedman) {\it Cambridge: Cambridge Univ. Press, 2004}, p. 196
\bibitem{} 36. Elgaroy O., Lahav O., Percival W.J. et al., {\it Phys.
Rev. Lett.} {\bf 89} 061301 (2002) 
\bibitem{} 37. Percival W.J., Sutherland W., Peacock J.A. et al.,
{\it Mon. Not. R. Astron. Soc.} {\bf 337} 1068 (2002)
\bibitem{} 38. Croom S.M., Smith R.J., Boyle B.J. et al.
{\it Mon. Not. R. Astron. Soc.} {\bf 349} 1397 (2004)
\bibitem{} 39. Jones D.H., Saunders W., Colless M. et al., 
{\it Mon. Not. R. Astron. Soc.} {\bf 355} 747 (2004) 
\bibitem{} 40. Reshetnikov V.P., {\it Poverhnostnaya Fotometriya Galaktik}
(Surface Photometry of Galaxies) (St.Petersburg: Izd. SPbGU, 2003)
\bibitem{} 41. Lynden-Bell D., Faber S.M., Burstein D. et al., 
{\it Astrophys. J.} {\bf 326} 19 (1988)
\bibitem{} 42. York D.G., Adelman J., Anderson J.E. et al., 
{\it Astron. J.} {\bf 120} 1579 (2000)
\bibitem{} 43. Stoughton Ch., Lupton R.H., Bernardi M. et al.,
{\it Astron. J.} {\bf 123} 485 (2002)
\bibitem{} 44. Abazajian K., Adelman-McCarthy J.K., Agueros M.A.
et al., {\it Astron. J.}, {\bf 129} 1755 (2005)
\bibitem{} 45. Loveday J., {\it Contemp. Phys.} {\bf 43} 437 (2002)  
\bibitem{} 46. Koo D.C., Kron R.G., {\it Ann. Rev. Astron. Astrophys.} 
{\bf 30} 613 (1992)
\bibitem{} 47. Sandage A., in {\it The Deep Universe} (Eds B Binggeli,
R Buser) {\it Berlin: Springer, 1995}, p.1
\bibitem{} 48. Ellis R.S., {\it Ann. Rev. Astron. Astrophys.} {\bf 35} 389
(1997)
\bibitem{} 49. Metcalfe N., Shanks T., Campos A. et al., 
{\it Mon. Not. R. Astron. Soc.} {\bf 323} 795 (2001)
\bibitem{} 50. McCracken H.J., Shanks T., Metcalfe N. et al., 
{\it Mon. Not. R. Astron. Soc.} {\bf 318} 913 (2000)
\bibitem{} 51. Ferguson H.C., Dickinson M., Williams R., 
{\it Ann. Rev. Astron. Astrophys.} {\bf 38} 667 (2000)
\bibitem{} 52. Williams R.E., Blacker B., Dickinson M. et al., {\it Astron. J.} 
{\bf 112} 1335 (1996)
\bibitem{} 53. Williams R.E., Baum S., Bergeron L.E. et al., {\it Astron. J.} 
{\bf 120} 2735 (2000)
\bibitem{} 54. Giacconi R., Zirm A., Wang J.X. et al., {\it Astrophys. J. Suppl.}
{\bf 139} 369 (2002)
\bibitem{} 55. Rosati P., Tozzi P., Giacconi R. et al., {\it Astrophys. J.}
{\bf 566} 667 (2002)
\bibitem{} 56. Alexander D.M., Bauer F.E., Brandt W.N. et al., {\it Astron. J.} 
{\bf 126} 539 (2003)
\bibitem{} 57. Barger A.J., Cowie L.L., Capak P. et al., {\it Astron. J.} 
{\bf 126} 632 (2003)
\bibitem{} 58. Lehmer B.D., Brandt W.N., Alexander D.M. et al., {\it Astron. J.} 
{\bf 129} 1 (2005)
\bibitem{} 59. Heidt J., Appenzeller I., Gabasch A. et al., {\it Astron.
Astrophys.} {\bf 398} 49 (2003)
\bibitem{} 60. Noll S., Mehlert D., Appenzeller I. et al., {\it Astron.
Astrophys.} {\bf 418} 885 (2004)
\bibitem{} 61. Appenzeller I., Bender R., Bohm A. et al., {\it ESO Messenger}
{\bf 116} 18 (2004)
\bibitem{} 62. Kashikawa N., Shimasaku K., Yasuda N. et al., {\it Publ.
Astron. Soc. Japan} {\bf 56} 1011 (2004)
\bibitem{} 63. Maihara T., Iwamuro F., Tanabe H. et al., {\it Publ.
Astron. Soc. Japan} {\bf 53} 25 (2001)
\bibitem{} 64. Taniguchi Y., Ajiki M., Nagao T. et al., {\it Publ.
Astron. Soc. Japan} {\bf 57} 165 (2005)
\bibitem{} 65. Kodama T., Yamada T., Akiyama M. et al., {\it Mon. Not. R. 
Astron. Soc.} {\bf 350} 1005 (2004)
\bibitem{} 66. Ouchi M., Shimasaku K., Akiyama M. et al., {\it Astrophys. J.}
{\bf 620} L1 (2005)
\bibitem{} 67. Wolf C., Dye S., Kleinheinrich M. et al., {\it Astron.
Astrophys.} {\bf 377} 442 (2001)
\bibitem{} 68. Wolf C., Meisenheimer K., Rix H.-W. et al., {\it Astron.
Astrophys.} {\bf 401} 73 (2003)
\bibitem{} 69. Kleinheinrich M., Schneider P., Rix H.-W. et al., {\it Astron.
Astrophys.} in press; astro-ph/0412615
\bibitem{} 70. Wolf C., Meisenheimer K., Kleinheinrich M. et al., {\it Astron.
Astrophys.} {\bf 421} 913 (2004)
\bibitem{} 71. Giavalisco M., Ferguson H.C., Koekemoer A.M. et al.,
{\it Astrophys. J.} {\bf 600} L93 (2004)
\bibitem{} 72. Riess A.G., Strolger L.-G., Tonry J. et al., {\it Astrophys. J.}
{\bf 607} 665 (2004)
\bibitem{} 73. Beckwith S., Somerville R., Stiavelli M., {\it STScI Newsletter}
{\bf 20} Issue 4 (2003)
\bibitem{} 74. Bouwens R.J., Thompson R.I., Illingworth G.D. et al.,
{\it Astrophys. J.} {\bf 616} L79 (2004)
\bibitem{} 75. Shectman S.A., Landy S.D., Oemler A. et al., {\it Astrophys. J.} 
{\bf 470} 172 (1996)
\bibitem{} 76. Arnouts S., D'Odorico S., Cristiani S. et al., {\it Astron.
Astrophys.} {\bf 341} 641 (1999)
\bibitem{} 77. Yee H.K.C., Morris S.L., Lin H. et al., {\it Astrophys. J.
Suppl.} {\bf 129} 475 (2000)
\bibitem{} 78. Drory N., Feulner G., Bender R. et al., {\it Mon. Not. R. 
Astron. Soc.} {\bf 325} 550 (2001)
\bibitem{} 79. Cimatti A., Mignoli M., Daddi E. et al., {\it Astron.
Astrophys.} {\bf 392} 395 (2002)
\bibitem{} 80. Davis M., Faber S.M., Newman J.A. et al., {\it Proc. SPIE}
{\bf 4834} 161 (2002); astro-ph/0209419
\bibitem{} 81. Rix H.-W., Barden M., Beckwith S. et al., {\it Astrophys. J.
Suppl.} {\bf 152} 163 (2004)
\bibitem{} 82. Le Fevre O., Mellier Y., McCracken H.J. et al.,
{\it Astron. Astrophys.} {\bf 417} 839 (2004)
\bibitem{} 83. Le Fevre O., Vettolani G., Garilli B. et al., {\it Astron. 
Astrophys.} (2005); astro-ph/0409133
\bibitem{} 84. Alcala J.M., Pannella M., Puddu E. et al. {\it Astron. 
Astrophys.} {\bf 428} 339 (2004)
\bibitem{} 85. Karachentsev I.D., {\it Pis'ma Astron. Zh.} 
{\bf 6} 3 (1980) [Sov. Astron. Lett. {\bf 6} 1 (1980)] 
\bibitem{} 86. Fatkhullin T.A., Vasil'ev A.A., Reshetnikov V.P. 
{\it Pis'ma Astron. Zh.} {\bf 30} 323 (2004) [Astron. Lett. {\bf 30} 
283 (2004)]
\bibitem{} 87. Sandage A., {\it Astrophys. J.} {\bf 133} 355 (1961)
\bibitem{} 88. Nagashima M., Yoshii Y., Totani T., Gouda N., 
{\it Astrophys. J.} {\bf 578} 675 (2002)
\bibitem{} 89. Gardner J.P., {\it Publ. Astron. Soc. Pacif.} {\bf 110}
291 (1998)
\bibitem{} 90. Peebles P.J.E. {\it The Large-Scale Structure of the 
Universe} (Princeton, NJ: Princeton University Press, 1980) 
\bibitem{} 91. Maller A.H., McIntosh D.H., Katz N., Weinberg M.D.,
{\it Astrophys. J.} {\bf 619} 147 (2005)
\bibitem{} 92. Jones B.J.T., Martinez V.J., Saar E., Trimble V.,
{\it Rev. Mod. Phys.} {\bf 76} 1211 (2004)
\bibitem{} 93. Budavari T., Connolly A.J., Szalay A.S. et al., 
{\it Astrophys. J.} {\bf 595} 59 (2003)
\bibitem{} 94. Maroto A.L., Ramirez J., astro-ph/0409280
\bibitem{} 95. Tegmark M., Blanton M.R., Strauss M.A. et al.,
{\it Astrophys. J.} {\bf 606} 702 (2004)
\bibitem{} 96. Kraan-Korteweg R.C., {\it Rev. Mod. Astron.}, in press;
astro-ph/0502217
\bibitem{} 97. Frith W.J., Busswell G.S., Fong R. et al., {\it Mon. Not. R. 
Astron. Soc.} {\bf 345} 1049 (2003)
\bibitem{} 98. Stiavelli M., Djorgovski S.G., Pavlovsky C. et al.,
{\it Astrophys. J.} {\bf 622} L1 (2005)
\bibitem{} 99. Miller L., Croom S.M., Boyle B.J. et al. {\it Mon. Not. R. 
Astron. Soc.} {\bf 355} 385 (2004)
\bibitem{} 100. Schechter P., {\it Astrophys. J.} {\bf 203} 297 (1976)
\bibitem{} 101. Norberg P., Cole Sh., Baugh C.M. et al., {\it Mon. Not. R. 
Astron. Soc.} {\bf 336} 907 (2002)
\bibitem{} 102. Croton D.J., Farrar G.R., Norberg P. et al. {\it Mon. Not. R. 
Astron. Soc.} {\bf 356} 1155 (2005)
\bibitem{} 103. Gabasch A., Bender R., Seitz S. et al., {\it Astron. 
Astrophys.} {\bf 421} 41 (2004)
\bibitem{} 104. Fried J.W., von Kuhlmann B., Meisenheimer K. et al.,
{\it Astron. Astrophys.} {\bf 367} 788 (2001)
\bibitem{} 105. Pozzetti L., Cimatti A., Zamorani G. et al., 
{\it Astron. Astrophys.} {\bf 402} 837 (2003)
\bibitem{} 106. Rudnick G., Rix H.-W., Franx M. et al. {\it Astrophys. J.} 
{\bf 599} 847 (2003) 
\bibitem{} 107. Reshetnikov V.P., Sotnikova N.Ya. {\it Astrofizika}
{\bf 36} 435 (1993) [Astrophysics {\bf 36} 265 (1994)]
\bibitem{} 108. Karachentsev I.D., Makarov D.I., in {\it Galaxy Interactions
at Low and High Redshift} (Eds J E Barnes, D B Sanders) 
{\it Dordrecht: Kluwer Acad. Publ., 1999}, p.109
\bibitem{} 109. Abraham R.G., Tanvir N.R., Santiago B.X. et al., 
{\it Mon. Not. R. Astron. Soc.} {\bf 279} L47 (1996)
\bibitem{} 110. Conselice Ch.J., Bershady  M.A., Jangren A., {\it Astrophys. J.} 
{\bf 529} 886 (2000)
\bibitem{} 111. Reshetnikov V.P., {\it Astron. Astrophys.} {\bf 353} 92 (2000)
\bibitem{} 112. Conselice Ch.J., Bershady M.A., Dickinson M., Papovich C.,
{\it Astron. J.} {\bf 126} 1183 (2003)
\bibitem{} 113. Hibbard J.E., Vacca W.D., {\it Astron. J.} {\bf 114} 1741 (1997)
\bibitem{} 114. Elmegreen B.G., Elmegreen D.M., Hirst A.C., {\it Astrophys. J.} 
{\bf 612} 191 (2004)
\bibitem{} 115. Jogee Sh., Barazza F.D., Rix H.-W. et al., {\it Astrophys. J.} 
{\bf 615} L105 (2004)
\bibitem{} 116. Lilly S., Schade D., Ellis R. et al., {\it Astrophys. J.} 
{\bf 500} 75 (1998)
\bibitem{} 117. Barden M., Rix H.-W., Somerville R.S. et al., 
{\it Astrophys. J.}, in press; astro-ph/0502416
\bibitem{} 118. Ziegler B.L., Bohm A., Fricke K.J. et al., {\it Astrophys. J.} 
{\bf 564} L69 (2002)
\bibitem{} 119. Reshetnikov V., Battaner E., Combes F., Jimenez-Vicente J.,
{\it Astron. Astrophys.} {\bf 382} 513 (2002)
\bibitem{} 120. Reshetnikov V.P., Dettmar R.-J., Combes F., 
{\it Astron. Astrophys.} {\bf 399} 879 (2003)
\bibitem{} 121. Kobulnicky H.A., Kewley L.J., {\it Astrophys. J.} {\bf 617} 
240 (2004) 
\bibitem{} 122. van der Wel A., Franx M., van Dokkum P.G., Rix H.-W.,
{\it Astrophys. J.} {\bf 601} L5 (2004)
\bibitem{} 123. Holden B.P., van der Wel A., Franx M. et al., 
{\it Astrophys. J.} {\bf 620} L83 (2005)
\bibitem{} 124. Treu T., in {\it Clusters of Galaxies: Probes of Cosmological
Structure and Galaxy Evolution} (Eds J S Mulchaey, A Dressler, A Oemler)
{\it Cambridge: Cambridge Univ. Press, 2004}, p.178
\bibitem{} 125. Stern D., Spinrad H., {\it Publ. Astron. Soc. Pacif.} {\bf 111}
1475 (1999)
\bibitem{} 126. Fan X., Strauss M.A., Schneider D.P. et al., {\it Astron. J.} 
{\bf 125} 1649 (2003)
\bibitem{} 127. Dey A., Spinrad H., Stern D. et al., {\it Astrophys. J.} 
{\bf 498} L93 (1998) 
\bibitem{} 128. Spinrad H., astro-ph/0308411
\bibitem{} 129. Reshetnikov V.P., Vasil'ev A.A. {\it Pis'ma Astron. Zh.}
{\bf 28} 3 (2002) [Astron. Lett. {\bf 28} 1 (2002)]
\bibitem{} 130. Ouchi M., Shimasaku K., Furusawa H. et al., {\it Astrophys. J.} 
{\bf 582} 60 (2003) 
\bibitem{} 131. Ouchi M., Shimasaku K., Okamura S. et al., {\it Astrophys. J.} 
{\bf 611} 685 (2004) 
\bibitem{} 132. Ferguson H.C., Dickinson M., Giavalisco M. et al.,
{\it Astrophys. J.} {\bf 600} L107 (2004) 
\bibitem{} 133. Bouwens R.J., Illingworth G.D., Blakeslee J.P. et al.,
{\it Astrophys. J.} {\bf 611} L1 (2004) 
\bibitem{} 134. Gunn J.E., Peterson B.A., {\it Astrophys. J.} {\bf 142} 
1633 (1965) 
\bibitem{} 135. Loeb A., Barkana R., {\it Ann. Rev. Astron. Astrophys.} 
{\bf 39} 19 (2001)
\bibitem{} 136. Becker R.H., Fan X., White R.L. et al., {\it Astron. J.} 
{\bf 122} 2850 (2001)
\bibitem{} 137. Sazhin M.V. {\it Usp. Fiz. Nauk} {\bf 174} 197 (2004)
[Phys. Usp. {\bf 47} 187 (2004)]
\bibitem{} 138. Wyithe J.S., Loeb A.  {\it Astrophys. J.} {\bf 586} 
693 (2003) 
\bibitem{} 139. Madau P., Ferguson H.C., Dickinson M.E. et al.
{\it Mon. Not. R. Astron. Soc.} {\bf 283} 1388 (1996)
\bibitem{} 140. Bouwens R.J., Illingworth G.D., Thompson R.I. et al.
{\it Astrophys. J.} {\bf 606} L25 (2004) 
\bibitem{} 141. Heavens A., Panter B., Jimenez R., Dunlop J. {\it Nature}
{\bf 428} 625 (2004)
\bibitem{} 142. Drory N., Salvato M., Gabasch A. et al. 
{\it Astrophys. J.} {\bf 619} L131 (2005) 
\bibitem{} 143. Gott J.R., Juric M., Schlegel D. et al. 
{\it Astrophys. J.} {\bf 624} 463 (2005) 
\bibitem{} 144. Chernin A.D. {\it Usp. Fiz. Nauk} {\bf 171} 1153 (2001)
[Phys. Usp. {\bf 44} 1099 (2001)]
\bibitem{} 145. Fall S.M., in {\it Building Galaxies: from the Primordial  
Universe to the Present} (Eds F Hammer, T X Thuan, V Cayatte et al.)
{\it XXX: World Scientific Publishing, 2000}, p.463
\bibitem{} 146. Djorgovski S.G., Williams R. astro-ph/0504006

\end{thebibliography}
\end{document}